\documentclass[twocolumn,showpacs,amsmath,amssymb,prd,floatfix]{revtex4}
\usepackage{graphicx}% Include figure files
\usepackage{dcolumn}% Align table columns on decimal point
\usepackage[varg]{txfonts}
\usepackage{bm}% bold math

\begin{document}
%\begin{CJK}{UTF8}{} % Use default fonts from CJK (see below)

\title{
%  Primordial Non-Gaussianity and Scale-dependent Bias in the
%  Integrated Perturbation Theory
  Deriving an Accurate Formula of Scale-dependent Bias with Primordial
  Non-Gaussianity:\\ An Application of the Integrated Perturbation
  Theory }

%\author{Takahiko Matsubara (松原隆彦)}
\author{Takahiko Matsubara} \email{taka@kmi.nagoya-u.ac.jp}
\affiliation{%
  Kobayashi-Maskawa Institute for the Origin of Particles and the
  Universe, Nagoya University, Chikusa, Nagoya, 464-8602, Japan; }%
\affiliation{%
  Department of Physics, Nagoya University, Chikusa, Nagoya, 464-8602,
  Japan}%

\date{\today}% It is always \today, today,

\begin{abstract}
    We apply the integrated perturbation theory \cite{mat11} to
    evaluate the scale-dependent bias in the presence of primordial
    non-Gaussianity. The integrated perturbation theory is a general
    framework of nonlinear perturbation theory, in which a broad class
    of bias models can be incorporated into perturbative evaluations
    of biased power spectrum and higher-order polyspectra.
    Approximations such as the high-peak limit or the peak-background
    split are not necessary to derive the scale-dependent bias in this
    framework. Applying the halo approach, previously known formulas
    are re-derived as limiting cases of a general formula in this
    work, and it is implied that modifications should be made in
    general situations. Effects of redshift-space distortions are
    straightforwardly incorporated. It is found that the slope of the
    scale-dependent bias on large scales is determined only by the
    behavior of primordial bispectrum in the squeezed limit, and is
    not sensitive to bias models in general. It is the amplitude of
    scale-dependent bias that is sensitive to the bias models. The
    effects of redshift-space distortions turn out to be quite small
    for the monopole component of the power spectrum, while the
    quadrupole component is proportional to the monopole component on
    large scales, and thus also sensitive to the primordial
    non-Gaussianity.
\end{abstract}

\maketitle

%\end{CJK}

\section{\label{sec:intro}
Introduction
}

The primordial non-Gaussianity is a useful indicator in searching for
the generation mechanism of density fluctuations in the universe.
While the primordial non-Gaussianity is small in the simplest model of
single-field slow-roll inflation, there are various other inflationary
models which predict fairly large primordial non-Gaussianities (for
review, see Refs.~\cite{BKMR04,che10}). Alternative models without
inflation can also produce large non-Gaussianities (see, e.g.,
Ref.~\cite{leh10}). Accordingly, constraining or detecting the
primordial non-Gaussianity has a substantial importance in studying
very early stages of the universe.

The large-scale structure (LSS) of the universe has been one of the
most important ways of constraining cosmological models in general. In
recent years, it was found that the primordial non-Gaussianity
produces the scale-dependent bias in the LSS \cite{dal08,MV08,DS10a}.
Although the scale-dependent bias from the primordial non-Gaussianity
mainly appears on very large scales, the form of the scale dependence
does not receive general relativistic corrections even on scales
larger than the Hubble radius \cite{WS09}.

The scale-dependent bias as a method to constrain the primordial
non-Gaussianity has already been applied to observational data of
galaxies and quasars \cite{slo08,xia10a,xia10b,xia11}. The constraints
derived from the scale-dependent bias are competitive with the
measurements in the cosmic microwave background (CMB). A hint of a
positive value of local-type non-Gaussianity parameter $f_{\rm NL}$
was indicated by analyses of the scale-dependent bias \cite{xia11},
which could have profound implications for models of the early
universe. However, more detailed analyses with large galaxy surveys
are necessary to derive conclusive results.

On the theoretical side, analytic expressions for the scale-dependent
bias has been only approximately derived. So far, at least three kinds
of derivation are known. One is based on the method of peak-background
split \cite{dal08,AT08,slo08,GP10,SK10,DJS11a,DJS11b,SHMC11}, and the
other is based on the statistics of high-threshold regions
\cite{MV08,VM09,JK09,DJS11b}. It is also shown that the
scale-dependent bias generally appears in phenomenological models of
local bias \cite{mcd08,TKM08}. Because those theoretical derivations
are approximate and not exact, they should have been compared with
numerical simulations \cite{DSI09,gro09,PPH10,WVB10,WV11}. The
scale-dependent bias is qualitatively confirmed by simulations,
although the detailed amplitude of the analytic predictions needs to
be modified \cite{DSI09,WV11,DJS11a}.

The purpose of this paper is to provide more precise and more general
analytic expressions for the scale-dependent bias in the presence of
primordial non-Gaussianity. Evolutions of density fluctuations on
sufficiently large scales are expected to be described by the
nonlinear perturbation theory. However, a consistent inclusion of the
general form of bias in the nonlinear perturbation theory had not been
clear until recently.

The integrated perturbation theory (iPT) \cite{mat11} is the formalism
in which a broad class of biasing scheme can be consistently included
on a general ground. The local Eulerian biasing scheme has been
frequently adopted in attempts of incorporating the bias into the
nonlinear perturbation theory. However, the local Eulerian bias is not
consistent with the nonlinear dynamics in general \cite{mat11,CSS12},
because the nonlinear evolutions are nonlocal phenomena. In the
formalism of iPT, generally nonlocal biasing either in Eulerian and
Lagrangian spaces can be consistently included in the nonlinear
perturbation theory. The effects of primordial non-Gaussianity and
redshift-space distortions are naturally incorporated into the
formalism.

Consequently, it is a straightforward application of iPT to making
predictions of scale-dependent bias in the presence of primordial
non-Gaussianity. In this paper, we first present the most general
prediction of iPT for the scale-dependent bias, which can be applied
to almost any type of primordial non-Gaussianity and to almost any
model of bias, as long as we consider the regime where the
perturbation theory applies. We then consider popular models of
primordial non-Gaussianity, i.e., local-, equilateral-, folded-, and
orthogonal-type non-Gaussianities. Asymptotic behaviors of
scale-dependent bias on large scales, which have been derived in
limited cases in the literature, are re-derived in general situations
without resorting to specific forms of bias. When the halo model of
bias is adopted, the previously known formulas of scale-dependent bias
are re-derived by taking appropriate limits of our general formula. In
course of derivation, nonlocality of bias turns out to be important.
We also show that previous formulas derived from the peak-background
split are only consistent with the Press-Schechter mass function. When
the mass function deviates from the Press-Schechter form, our general
formula predicts that those previous formulas of scale-dependent bias
should be corrected.

The main purpose of this paper is to show how the iPT can be applied
to making theoretical predictions of the scale-dependent bias, and to
give lowest-order calculations with primordial bispectra, when the
bias is given by a simple, nonlocal model of halo bias. With the iPT
formalism, predicting the scale-dependent bias is straightforward once
the bias model is given. Theoretical uncertainties in predicting the
scale-dependent bias are reduced to those of the bias model. Accurate
modeling of biasing is actively studied in recent years. Once we have
an accurate model of galaxy bias which is generally nonlocal, the iPT
immediately gives predictions of scale-dependent bias with least
approximations.

This paper is organized as follows. In Sections
\ref{sec:SDBias}--\ref{sec:BiasFn}, analytic derivations of
scale-dependent bias in real space are presented in order. The general
formula of the biased power spectrum in real space with primordial
bispectrum is derived by the lowest-order iPT in
Sec.~\ref{sec:SDBias}. Large-scale limits of the scale-dependent bias
in concrete models of primordial non-Gaussianity are generally
investigated in Sec.~\ref{sec:PNGmodels} without assuming the models
of bias. In Sec.~\ref{sec:BiasFn}, we consider the shapes of
renormalized bias functions which are needed in iPT. We generalize the
previous results to include the effects of the smoothing function in
the halo model of bias. Numerical evaluations of the analytic formula
are presented and compared with previous predictions in
Sec.~\ref{sec:Numerical}. In Sec.~\ref{sec:SDBiasR}, we generalize our
formula to include the redshift-space distortions. In
Sec.~\ref{sec:concl}, we summarize our results.

In plotting the figures of this paper, we adopt a cosmological model
of flat curvature with parameters $\varOmega_{\rm m0} = 0.275$,
$\varOmega_{\Lambda 0} = 1-\varOmega_{\rm m0} = 0.725$,
$\varOmega_{\rm b0} = \varOmega_{\rm m0} - \varOmega_{\rm c0} =
0.046$, $n_{\rm s} = 0.96$, $\sigma_8 = 0.8$, $H_0 = 70\,{\rm
  km/s/Mpc}$, where $\varOmega_{\rm m0}$ is the matter density
parameter, $\varOmega_{\Lambda 0}$ is the cosmological constant
parameter, $\varOmega_{\rm b0}$ is the baryon density parameter,
$\varOmega_{\rm c0}$ is the cold dark matter density parameter,
$n_{\rm s}$ and $\sigma_8$ are respectively the spectral index and the
amplitude of primordial density fluctuations, and $H_0$ is the
Hubble's constant.

\section{\label{sec:SDBias} Integrated perturbation theory and
  scale-dependent bias from primordial non-Gaussianity}

We apply the formalism of iPT to derive a general formula of the
scale-dependent bias from the primordial non-Gaussianity, without
assuming the peak-background split or the high-peak limit. In the
derivation, it is convenient to introduce the multipoint propagators
of biased objects. The multipoint propagators of mass and velocity
fields are recently introduced in Ref.~\cite{BCS08,BCS10}, and those
of biased objects are introduced in Ref.~\cite{mat11}. The multipoint
propagators $\varGamma_{\rm X}^{(n)}$ of biased objects X are defined
by ensemble averages of functional derivatives
\begin{equation}
  \left\langle
      \frac{\delta^n \delta_{\rm X}(\bm{k})}
      {\delta\delta_{\rm L}(\bm{k}_1)
        \cdots\delta\delta_{\rm L}(\bm{k}_n)}
  \right\rangle =
  (2\pi)^{3-3n}\delta_{\rm D}^3(\bm{k}_{1\cdots n}-\bm{k})
  \varGamma^{(n)}_{\rm X}(\bm{k}_1,\ldots,\bm{k}_n),
\label{eq:2-1}
\end{equation}
where $\delta_{\rm L}(\bm{k})$ is the Fourier transform of the linear
density field, and $\delta_{\rm X}(\bm{k})$ is the Fourier transform
of the number density field of the biased objects X {\em in Eulerian
  space}. On the right-hand side, $\bm{k}_{1\cdots n} \equiv
\bm{k}_1+\cdots +\bm{k}_n$ and $\delta_{\rm D}^3(\bm{k})$ is the
Dirac's delta function in three dimensions.

\begin{figure}
\begin{center}
\includegraphics[width=20.5pc]{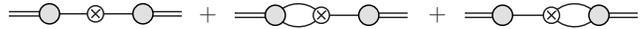}
\caption{\label{fig:ps}
The diagrammatic representation of the power spectrum in terms of
multipoint propagators. Details of diagrammatic rules are described in
Ref.~\cite{mat11}.
}
\end{center}
\end{figure}

In terms of the multipoint propagators, the biased power spectrum of
objects X is represented by
\begin{align}
  P_{\rm X}(k) &= [\varGamma_{\rm X}^{(1)}(\bm{k})]^2 P_{\rm L}(k) 
\nonumber\\
  & \quad + \varGamma_{\rm X}^{(1)}(\bm{k})
  \int \frac{d^3k'}{(2\pi)^3}
  \varGamma_{\rm X}^{(2)}(\bm{k}',\bm{k}-\bm{k}')
  B_{\rm L}(k,k',|\bm{k}-\bm{k}'|)
\nonumber\\
  &\quad + \cdots,
\label{eq:2-2}
\end{align}
where $P_{\rm L}(k)$ is the linear power spectrum and $B_{\rm
  L}(k_1,k_2,k_3)$ is the linear bispectrum. The diagrammatic
representation of the above equation is shown in Fig.~\ref{fig:ps}.
The details of the diagrammatic rules are described in
Ref.~\cite{mat11}.
The multipoint propagators are represented by grey circles and the
linear polyspectra are represented by crossed circles. The multipoint
propagators include all the loop corrections at the corresponding
vertices. For Gaussian initial conditions, the linear bispectrum is
absent. In this paper, we consider only lowest-order contributions of
the primordial non-Gaussianity which are linearly dependent on the
bispectrum. Contributions from the trispectrum and higher-order
polyspectra, and terms like a product of power spectrum and bispectrum
are all neglected.

The multipoint propagators $\varGamma_{\rm X}^{(1)}$ and
$\varGamma_{\rm X}^{(2)}$ also contain contributions from primordial
non-Gaussianity in general. However, such contributions in
Eq.~(\ref{eq:2-2}) are higher orders in the above sense. Accordingly,
we do not have to consider the non-Gaussian corrections to evaluate
the multipoint propagators of Eq.~(\ref{eq:2-1}) in the lowest-order
approximation of this paper.

In the iPT, a concept of renormalizing Lagrangian bias functions is
introduced. The renormalized bias functions in Lagrangian space
$c^{\rm L}_n$ is defined by
\begin{equation}
  c^{\rm L}_n(\bm{k}_1,\ldots,\bm{k}_n) =
  (2\pi)^{3n} \int \frac{d^3k}{(2\pi)^3}
  \left\langle
      \frac{\delta^n \delta^{\rm L}_{\rm X}(\bm{k})}
      {\delta\delta_{\rm L}(\bm{k}_1)
        \cdots\delta\delta_{\rm L}(\bm{k}_n)}
  \right\rangle,
\label{eq:2-3}
\end{equation}
where $\delta^{\rm L}_{\rm X}(\bm{k})$ is the Fourier transform of the
number density field of biased objects X {\em in Lagrangian space}.
An equivalent way of defining the renormalized bias functions is given
by
\begin{equation}
  \left\langle
      \frac{\delta^n \delta^{\rm L}_{\rm X}(\bm{k})}
      {\delta\delta_{\rm L}(\bm{k}_1)
        \cdots\delta\delta_{\rm L}(\bm{k}_n)}
  \right\rangle = 
  (2\pi)^{3-3n}\delta_{\rm D}^3(\bm{k}_{1\cdots n}-\bm{k})
  c^{\rm L}_n(\bm{k}_1,\ldots,\bm{k}_n).
\label{eq:2-4}
\end{equation}
In this form of definition, the similarity of the renormalized bias
functions in Lagrangian space and the multipoint propagators of
Eulerian space is obvious. The renormalized bias functions $c^{\rm
  L}_n$ can be seen as multipoint propagators of biasing in Lagrangian
space. When the bias is not present, the number density field of
objects is uniform in Lagrangian space, $\delta^{\rm L}_{\rm X} = 0$,
and all the renormalized bias functions vanish.

\begin{figure}
\begin{center}
\includegraphics[width=20.5pc]{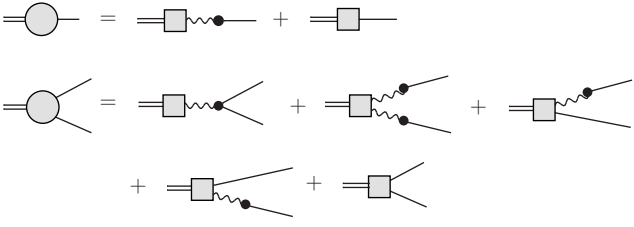}
\caption{\label{fig:multp} Diagrammatic representations of the first
  two multipoint propagators with renormalized Lagrangian bias
  functions in the lowest-order approximation. Details of diagrammatic
  rules are described in Ref.~\cite{mat11}. }
\end{center}
\end{figure}

Applying the diagrammatic methods of iPT \cite{mat11} in the
lowest-order approximation, the first and second multipoint
propagators in terms of renormalized Lagrangian bias functions are
given by
\begin{align}
    \varGamma_{\rm X}^{(1)}(\bm{k}) &= \bm{k}\cdot\bm{L}_1(\bm{k}) +
    c^{\rm L}_1(k)
\label{eq:2-4-1a}\\
  \varGamma_{\rm X}^{(2)}(\bm{k}_1,\bm{k}_2)
  &= \bm{k}\cdot\bm{L}_2(\bm{k}_1,\bm{k}_2)
  + \left[\bm{k}\cdot\bm{L}_1(\bm{k}_1)\right]
    \left[\bm{k}\cdot\bm{L}_1(\bm{k}_2)\right]
\nonumber\\
& \quad
  + \bm{k}\cdot\bm{L}_1(\bm{k}_1) c^{\rm L}_1(\bm{k}_2)
  + \bm{k}\cdot\bm{L}_1(\bm{k}_2) c^{\rm L}_1(\bm{k}_1)
\nonumber\\
& \quad
  + c^{\rm L}_2(\bm{k}_1,\bm{k}_2),
\label{eq:2-4-1b}
\end{align}
where $\bm{L}_n$ is the $n$th order perturbation kernel of
displacement field, and $\bm{k}=\bm{k}_1 + \bm{k}_2$ in
Eq.~(\ref{eq:2-4-1b}). The diagrams for the above expressions are
shown in Fig.~\ref{fig:multp}.
The correspondences between diagrams and equations are detailed in
Ref.~\cite{mat11}. We omit partial resummations of the displacement
field $\bm{\varPsi}$. If we leave these partial resummations, the
right-hand sides of Eqs.~(\ref{eq:2-4-1a}) and (\ref{eq:2-4-1b}) are
multiplied by a factor $\Pi(\bm{k}) = \langle
e^{-i\bm{k}\cdot\bm{\varPsi}}\rangle$. This factor is almost unity on
sufficiently large scales. Moreover, this factor is cancelled when we
calculate the scale-dependent bias $\varDelta b(k)$ of
Eq.~(\ref{eq:2-11}) below. Therefore, we do not include this factor
from the first place. One should note that Eqs.~(\ref{eq:2-4-1a}) and
(\ref{eq:2-4-1b}) are lowest-order results. For higher-order
calculations, loop corrections to the tree-level expressions of
Eqs.~(\ref{eq:2-4-1a}) and (\ref{eq:2-4-1b}) for multipoint
propagators should be included as depicted in Fig.~17 of
Ref.~\cite{mat11}.

The perturbation kernels $\bm{L}_n$ are calculated by the Lagrangian
perturbation theory \cite{buc89,mou92,ber94,cat95,CT96,mat08a,RB12}.
In real space, the first two kernels are given by
\begin{align}
  \bm{L}_1(\bm{k}) &= \frac{\bm{k}}{k^2},
  \label{eq:2-4-2a}\\
  \bm{L}_2(\bm{k}_1,\bm{k}_2) &=
  \frac{3}{7} \frac{\bm{k}_{12}}{{k_{12}}^2}
  \left[1 - \left(\frac{\bm{k}_1\cdot\bm{k}_2}{k_1k_2}\right)^2\right],
\label{eq:2-4-2b}
\end{align}
and Eqs.~(\ref{eq:2-4-1a}) and (\ref{eq:2-4-1b}) reduce to
\begin{align}
    \varGamma_{\rm X}^{(1)}(\bm{k}) &= 1 + c^{\rm L}_1(k)
\label{eq:2-4a}\\
  \varGamma_{\rm X}^{(2)}(\bm{k}_1,\bm{k}_2) &= F_2(\bm{k}_1,\bm{k}_2)
  + \left(1+\frac{\bm{k}_1\cdot\bm{k}_2}{{k_1}^2}\right) c^{\rm L}_1(k_2)
\nonumber\\
& \quad
  + \left(1+\frac{\bm{k}_1\cdot\bm{k}_2}{{k_2}^2}\right) c^{\rm L}_1(k_1)
  + c^{\rm L}_2(\bm{k}_1,\bm{k}_2)
\label{eq:2-4b}
\end{align}
where
\begin{equation}
  F_2(\bm{k}_1,\bm{k}_2) = \frac{10}{7} + 
  \left(\frac{k_1}{k_2} +\frac{k_2}{k_1} \right)
  \frac{\bm{k}_1\cdot\bm{k}_2}{k_1k_2}
  + \frac{4}{7}\left(\frac{\bm{k}_1\cdot\bm{k}_2}{k_1k_2}\right)^2,
\label{eq:2-5}
\end{equation}
is the second-order kernel of the Eulerian perturbation theory
\cite{BCGS02}. In the literature, an extra factor of $1/2!$ is
frequently multiplied to the right-hand side of Eq.~(\ref{eq:2-5}). We
find it is convenient not to include the factor for symmetrical
reasons in diagrammatic methods \cite{mat11}.

In the lowest-order approximation, non-Gaussian contributions to the
renormalized bias functions can be neglected as described earlier.
Consequently, we can consider that the first term of right-hand side
of Eq.~(\ref{eq:2-2}) corresponds to the Gaussian part $P_{\rm X}^{\rm
  G}(k)$ and the second term corresponds to the non-Gaussian part
$\varDelta P_{\rm X}(k)$:
\begin{align}
  P_{\rm X}^{\rm G}(k) &= [\varGamma_{\rm X}^{(1)}(\bm{k})]^2 P_{\rm L}(k),
\label{eq:2-5a}\\
  \varDelta P_{\rm X}(k) &= \varGamma_{\rm X}^{(1)}(\bm{k})
  \int \frac{d^3k_1}{(2\pi)^3} \frac{d^3k_2}{(2\pi)^3}
  (2\pi)^3\delta_{\rm D}(\bm{k}_1+\bm{k}_2-\bm{k})
\nonumber\\
  &\hspace{6.5pc} \times
  \varGamma_{\rm X}^{(2)}(\bm{k}_1,\bm{k}_2)
  B_{\rm L}(k,k_1,k_2).
\label{eq:2-5b}
\end{align}
One should note that the above distinction holds only in the
lowest-order approximation. In general, there are non-Gaussian
corrections to the first multipoint propagator $\varGamma^{(1)}_{\rm
  X}$ beyond the lowest-order approximation. Those corrections in
Eq.~(\ref{eq:2-5a}) should be included in $\varDelta P_{\rm X}(k)$ if
we consider the higher-order approximations. For example, the
lowest-order non-Gaussian correction term is proportional to the
primordial bispectrum. However, such correction terms are multiplied
by the power spectrum in Eq.~(\ref{eq:2-5a}). Accordingly, the orders
of non-Gaussian corrections to the multipoint propagator in
Eq.~(\ref{eq:2-5a}) are higher than those of Eq.~(\ref{eq:2-5b}), and
can be neglected in the lowest-order approximation. For the same
reason, we do not consider loop corrections to the Gaussian part
$P_{\rm X}^{\rm G}(k)$ which are extensively studied in previous
papers \cite{mat08a,mat08b,oka11,sat11} with Gaussian initial
conditions. Evaluations of the leading non-Gaussian part $\varDelta
P_{\rm X}(k)$ is the main subject in this paper.

Substituting Eqs.~(\ref{eq:2-4a}) and (\ref{eq:2-4b}) into
Eqs.~(\ref{eq:2-5a}) and (\ref{eq:2-5b}) we have
\begin{align}
  P^{\rm G}_{\rm X}(k) &= \left[b_1(k)\right]^2
  P_{\rm L}(k),
\label{eq:2-6a}\\
  \varDelta P_{\rm X}(k)
  &= b_1(k)
  \left[Q_0(k) + Q_1(k) + Q_2(k) 
  \right],
\label{eq:2-6b}
\end{align}
where
\begin{equation}
  b_1(k) = 1 + c^{\rm L}_1(k)
\label{eq:2-7}
\end{equation}
corresponds to the Eulerian linear bias factor, and the functions
$Q_n(k)$ are defined by
\begin{equation}
  Q_n(k) = \int \frac{d^3k'}{(2\pi)^3}
  \hat{Q}_n(\bm{k},\bm{k}') B_{\rm L}(k,k',|\bm{k}-\bm{k'}|),
\label{eq:2-8}
\end{equation}
where
\begin{align}
  \hat{Q}_0(\bm{k},\bm{k}') &=
  \frac{2\bm{k}\cdot\bm{k}'}{k'^2}
  -\frac47 \frac{k^2k'^2 - (\bm{k}\cdot\bm{k}')^2}{k'^2|\bm{k}-\bm{k}'|^2},
\label{eq:2-9a}\\
  \hat{Q}_1(\bm{k},\bm{k}') &=
  \frac{2\bm{k}\cdot\bm{k}'}{k'^2}
  c^{\rm L}_1(|\bm{k}-\bm{k}'|),
\label{eq:2-9b}\\
  \hat{Q}_2(\bm{k},\bm{k}')
  &= c^{\rm L}_2\left(\bm{k}',\bm{k}-\bm{k}'\right).
\label{eq:2-9c}
\end{align}
When the bias is local, the linear bias factor $b_1(k)$ is
scale-independent. In general, the scale-dependent bias emerges from
nonlocality even when the primordial non-Gaussianity is absent.

When the scale-dependent bias $\varDelta b(k)$ from the primordial
non-Gaussianity is defined by
\begin{equation}
  P_{\rm X}(k) = \left[b_1(k) + \varDelta b(k)\right]^2 P_{\rm m}(k),
\label{eq:2-10}
\end{equation}
and higher orders of $\varDelta b$ are neglected, we have
\begin{equation}
  \varDelta b(k) = \frac12 b_1(k)
  \left(
      \frac{\varDelta P_{\rm X}(k)}{P_{\rm X}^{\rm G}(k)}
    - \frac{\varDelta P_{\rm m}(k)}{P_{\rm m}^{\rm G}(k)}
  \right),
\label{eq:2-11}
\end{equation}
where $P_{\rm m}^{\rm G}(k)$ is the power spectrum of mass without
primordial non-Gaussianity and $\varDelta P_{\rm m}(k)$ is the
contribution from the primordial non-Gaussianity, so that $P_{\rm
  m}(k) = P_{\rm m}^{\rm G}(k) + \varDelta P_{\rm m}(k)$ is the total
power spectrum of mass. These are given by substituting $c^{\rm L}_1 =
c^{\rm L}_2 = 0$ ($b_1=1$) in the expressions of $P^{\rm G}_{\rm
  X}(k)$ and $\varDelta P_{\rm X}(k)$, respectively:
\begin{align}
  P^{\rm G}_{\rm m}(k) &= P_{\rm L}(k),
\label{eq:2-12a}\\
  \varDelta P_{\rm m}(k)
  &= Q_0(k).
\label{eq:2-12b}
\end{align}
Substituting Eqs.~(\ref{eq:2-6a}), (\ref{eq:2-6b}), (\ref{eq:2-12a})
and (\ref{eq:2-12b}) into Eq.~(\ref{eq:2-11}), we have
\begin{equation}
  \varDelta b(k) =
  \frac{1}{2P_{\rm L}(k)}
  \left[
      Q_1(k)  + Q_2(k) 
      -c^{\rm L}_1(k) Q_0(k)
  \right].
\label{eq:2-13}
\end{equation}

The above expression for the scale-dependent bias is consistent to
that derived from the cross power spectrum $P_{\rm mX}(k)$ between
mass and objects X as shown below. In fact, the cross power spectrum
is given by
\begin{align}
  P_{\rm mX}(k) &= \varGamma_{\rm m}^{(1)}(\bm{k})
  \varGamma_{\rm X}^{(1)}(\bm{k}) P_{\rm L}(k) 
\nonumber\\
  & + \frac12 \varGamma_{\rm m}^{(1)}(\bm{k})
  \int \frac{d^3k'}{(2\pi)^3}
  \varGamma_{\rm X}^{(2)}(\bm{k}',\bm{k}-\bm{k}')
  B_{\rm L}(k,k',|\bm{k}-\bm{k}'|)
\nonumber\\
  & + \frac12 \varGamma_{\rm X}^{(1)}(\bm{k})
  \int \frac{d^3k'}{(2\pi)^3}
  \varGamma_{\rm m}^{(2)}(\bm{k}',\bm{k}-\bm{k}')
  B_{\rm L}(k,k',|\bm{k}-\bm{k}'|)
\nonumber\\
  & + \cdots,
\label{eq:2-14}
\end{align}
where $\varGamma^{(n)}_{\rm m}$ is the $n$th multipoint propagator of
mass. In the lowest-order approximation, the first two propagators are
obtained by substituting $c^{\rm L}_1 = c^{\rm L}_2 = 0$ in
Eqs.~(\ref{eq:2-4a}) and (\ref{eq:2-4b}):
\begin{align}
  \varGamma_{\rm m}^{(1)}(\bm{k}) &= 1,
\label{eq:2-15a}\\
  \varGamma_{\rm m}^{(2)}(\bm{k}_1,\bm{k}_2) &= F_2(\bm{k}_1,\bm{k}_2).
\label{eq:2-15b}
\end{align}
Substituting Eqs.~(\ref{eq:2-4a}), (\ref{eq:2-4b}), (\ref{eq:2-15a})
and (\ref{eq:2-15b}) into Eq.~(\ref{eq:2-14}), we have
\begin{align}
  P^{\rm G}_{\rm mX}(k) &= b_1(k) P_{\rm L}(k),
\label{eq:2-16a}\\
  \varDelta P_{\rm mX}(k)
  &= \frac{1 + b_1(k)}{2} Q_0(k) + \frac12 Q_1(k) + \frac12 Q_2(k).
\label{eq:2-16b}
\end{align}
When the scale-dependent bias $\varDelta b(k)$ is defined by
\begin{equation}
  P_{\rm mX}(k) = \left[b_1(k) + \varDelta b(k)\right] P_{\rm m}(k),
\label{eq:2-17}
\end{equation}
and higher orders of $\varDelta b$ are neglected, we have
\begin{equation}
  \varDelta b(k) = b_1(k)
  \left(
      \frac{\varDelta P_{\rm mX}(k)}{P_{\rm mX}^{\rm G}(k)}
    - \frac{\varDelta P_{\rm m}(k)}{P_{\rm m}^{\rm G}(k)}
  \right).
\label{eq:2-18}
\end{equation}
Substituting Eqs.~(\ref{eq:2-12a}), (\ref{eq:2-12b}), (\ref{eq:2-16a})
and (\ref{eq:2-16b}) into Eq.~(\ref{eq:2-18}), we again have the same
equation as Eq.~(\ref{eq:2-13}).

Our result, Eq.~(\ref{eq:2-13}), is a general formula and any
approximation other than the perturbation theory is not employed. In
the literature, analytic formulas of scale-dependent bias from the
primordial non-Gaussianity have been derived either in the high-peak
limit or in the approximation of peak-background split. These
approximate results are subclasses of our general formula, as we
explicitly show in the rest of this paper.

\section{\label{sec:PNGmodels}
Models of primordial non-Gaussianity}

\subsection{\label{subsec:bispectra}
Specific models of primordial bispectra}

Models of primordial non-Gaussianity are quite commonly characterized
by the primordial bispectra $B_\varPhi$ of gauge-invariant Newtonian
potential $\varPhi$ at the matter-dominated epoch. The linear density
contrast $\delta_{\rm L}$ is proportional to $\varPhi$ in Fourier
space, and we have
\begin{equation}
  \delta_{\rm L}(\bm{k}) = {\cal M}(k) \varPhi(\bm{k}).
\label{eq:3-1}
\end{equation}
Throughout this paper, we do not explicitly write an argument of time
or redshift $z$ for simplicity in some time-dependent variables. In
the above equation, $\delta_{\rm L} \propto D(z)$ and ${\cal M}
\propto D(z)$ where $D(z)$ is the linear growth factor. However, the
potential $\varPhi$ is the primordial one and does not depend on $z$
by definition. Do not confuse with the physical Newtonian potential
which does depend on $z$. The proportional factor ${\cal M}(k)$ is
determined by the transfer function $T(k)$ and the Poisson equation
as
\begin{equation}
  {\cal M}(k) = \frac23
  \frac{D(z)}{(1 + z_*) D(z_*)}
  \frac{k^2T(k)}{{H_0}^2 \varOmega_{\rm m0}},
\label{eq:3-2}
\end{equation}
where $z_*$ is an arbitrary redshift at the matter-dominated epoch.
The factor $(1+z_*)D(z_*)$ does not depend on the choice of $z_*$ as
long as $z_*$ is deep in the matter-dominated epoch, since $D(z_*)
\propto 1/(1+z_*)$ in that epoch. Some authors employ the
normalization of the growth factor as $(1+z_*)D(z_*) = 1$, in which
case Eq.~(\ref{eq:3-2}) has the simplest form.

The linear power spectrum and bispectrum of $\delta_{\rm L}$ are given
by
\begin{align}
  P_{\rm L}(k) &= {\cal M}^2(k) P_\varPhi(k)
\label{eq:3-3a}\\
  B_{\rm L}(k_1,k_2,k_3) &= 
  {\cal M}(k_1) {\cal M}(k_2) {\cal M}(k_3) 
  B_\varPhi(k_1,k_2,k_3),
\label{eq:3-3b}
\end{align}
where $P_\varPhi(k)$ and $B_\varPhi(k_1,k_2,k_3)$ are the primordial
power spectrum and bispectrum of the potential, respectively. In
popular models of non-Gaussianity, the primordial bispectra are
uniquely related to the shape of primordial power spectrum.

There are four models of primordial bispectra which are frequently
considered as typical examples. Defining
\begin{equation}
  P_i \equiv P_\varPhi(k_i),
\label{eq:3-4}
\end{equation}
the four models are given by the following equations:
\begin{itemize}
\item
the {\em local} model \cite{GLMM94,VWHK00,KS01}:
\begin{equation}
  B_\varPhi^{\rm loc.}(k_1,k_2,k_3) = 2 f_{\rm NL}
  \left[ P_1 P_2 + \mbox{cyc.}\right].
\label{eq:3-5}
\end{equation}
\item
the {\em equilateral} model \cite{cre06}:
\begin{multline}
  B_\varPhi^{\rm eql.}(k_1,k_2,k_3) = 6 f_{\rm NL}
  \left[ -\,(P_1 P_2 + \mbox{cyc.}) - 2(P_1 P_2 P_3)^{2/3}
  \right.
\\ \left.
    +\, ({P_1}^{1/3}{P_2}^{2/3}P_3 + \mbox{5\ perm.}) \right].
\label{eq:3-6}
\end{multline}
\item
the {\em folded} model \cite{MSS09}:
\begin{multline}
  B_\varPhi^{\rm fol.}(k_1,k_2,k_3) = 6 f_{\rm NL}
  \left[ (P_1 P_2 + \mbox{cyc.}) + 3(P_1 P_2 P_3)^{2/3}
  \right.
\\ \left.
    -\, ({P_1}^{1/3}{P_2}^{2/3}P_3 + \mbox{5\ perm.}) \right].
\label{eq:3-7}
\end{multline}
\item
the {\em orthogonal} model \cite{SSZ10}:
\begin{multline}
  B_\varPhi^{\rm ort.}(k_1,k_2,k_3) = 6 f_{\rm NL}
  \left[-3(P_1 P_2 + \mbox{cyc.}) - 8(P_1 P_2 P_3)^{2/3}
  \right.
\\ \left.
    +3 ({P_1}^{1/3}{P_2}^{2/3}P_3 + \mbox{5\ perm.}) \right].
\label{eq:3-8}
\end{multline}
\end{itemize}
With these typical models of primordial bispectrum, the linear
bispectrum $B_{\rm L}(k_1,k_2,k_3)$ is given by the linear power
spectrum $P_{\rm L}(k)$ and a function ${\cal M}(k)$ through
Eqs.~(\ref{eq:3-3a}) and (\ref{eq:3-3b}). Once a model of the
primordial non-Gaussianity is given, the non-Gaussian part of the
power spectrum $\varDelta P_{\rm X}(k)$, $\varDelta P_{\rm mX}(k)$,
$\varDelta P_{\rm m}(k)$ of Eqs.~(\ref{eq:2-6b}), (\ref{eq:2-16b}),
(\ref{eq:2-12b}) and the scale-dependent bias of Eq.~(\ref{eq:2-13})
are straightforwardly evaluated for a given model of bias functions.

\subsection{\label{subsec:SqueezedLimit}
Large-scale limit of scale-dependent bias
}

In the large-scale limit $k \rightarrow 0$ of Eq.~(\ref{eq:2-8}), we
have $\hat{Q}_0(\bm{k},\bm{k}') \approx \hat{Q}_1(\bm{k},\bm{k}')
\approx 0$ in Eqs.~(\ref{eq:2-9a}) and (\ref{eq:2-9b}). Accordingly,
we have $Q_0(k) \approx Q_1(k) \approx 0$, and the scale-dependent
bias of Eq.~(\ref{eq:2-13}) approximately reduces to
\begin{equation}
  \varDelta b(k) \approx
  \frac{Q_2(k)}{2P_{\rm L}(k)}.
\label{eq:3-20}
\end{equation}
As long as the second-order renormalized bias function $c^{\rm
  L}_2(\bm{k}_1,\bm{k}_2)$ is a smooth function, we can adopt an
approximation
\begin{equation}
  c^{\rm L}_2(\bm{k}',\bm{k}-\bm{k}')
  \approx c^{\rm L}_2(\bm{k}',-\bm{k}')
\label{eq:3-20-1}
\end{equation}
in the large-scale limit for the integrand of $Q_2(k)$. Because of
rotational symmetry, $c^{\rm L}_2(\bm{k}',-\bm{k}')$ only depends on
the magnitude $k'=|\bm{k}'|$, and we define
\begin{equation}
  \tilde{c}^{\rm L}_2(k') \equiv c^{\rm L}_2(\bm{k}',-\bm{k}').
\label{eq:3-20-2}
\end{equation}
Assuming Eq.~(\ref{eq:3-20-1}) for the renormalized bias function, the
scale-dependent bias on large scales is asymptotically given by
\begin{equation}
  \varDelta b(k)
  \approx
  \frac{1}{2P_{\rm L}(k)} \int\frac{d^3k'}{(2\pi)^3}
  \tilde{c}^{\rm L}_2(k')
  B_{\rm L}\left(k,k',|\bm{k}-\bm{k}'|\right).
\label{eq:3-21}
\end{equation}
In this case, the scale dependence of $\varDelta b(k)$ is determined
by the functional form of the primordial bispectrum $B_{\rm
  L}(k,k',|\bm{k}-\bm{k}'|)$ in the squeezed limit $k \ll k'$.

For specific models of primordial non-Gaussianity introduced in the
previous subsection, squeezed limits of the bispectra can be
analytically derived. In the following, we assume the power-law
primordial spectrum, $P_{\varPhi}(k) \propto k^{n_{\rm s}-4}$, where
$n_{\rm s} \approx 1$ is the scalar spectral index. We define
\begin{equation}
  \alpha_{\rm s} = \frac{1-n_{\rm s}}{3},
\label{eq:3-22}
\end{equation}
to represent the deviation from the scale-free Harrison-Zel'dovich
spectrum, $n_{\rm s}=1$. Taking the squeezed limit $k \ll k'$ in
Eqs.~(\ref{eq:3-5})--(\ref{eq:3-8}), and keeping leading orders in
$k/k'$, they are asymptotically given by
\begin{align}
  B_\varPhi^{\rm loc.}(k,k',|\bm{k}-\bm{k}'|) &\approx
  4 f_{\rm NL} P_\varPhi(k) P_\varPhi(k'),
\label{eq:3-22a}\\
  B_\varPhi^{\rm eql.}(k,k',|\bm{k}-\bm{k}'|) &\approx
  12 f_{\rm NL}
  \left(\frac{k}{k'}\right)^2
  \left[
     \left(\frac{k}{k'}\right)^{2\alpha_{\rm s}}
     - (1 + \alpha_{\rm s})^2 \mu^2
  \right]
\nonumber\\
  & \qquad \times P_\varPhi(k) P_\varPhi(k'),
\label{eq:3-22b}\\
  B_\varPhi^{\rm fol.}(k,k',|\bm{k}-\bm{k}'|) &\approx
  6 f_{\rm NL}
  \frac{k}{k'} P_\varPhi(k) P_\varPhi(k'),
\label{eq:3-22c}\\
  B_\varPhi^{\rm ort.}(k,k',|\bm{k}-\bm{k}'|) &\approx
  -12 f_{\rm NL}
  \frac{k}{k'} P_\varPhi(k) P_\varPhi(k'),
\label{eq:3-22d}
\end{align}
where $\mu \equiv \bm{k}\cdot\bm{k}'/kk'$ is the direction cosine
between wavevectors $\bm{k}$ and $\bm{k}'$. Except for the equilateral
non-Gaussianity, the lowest-order bispectra divided by
$P_\varPhi(k)P_\varPhi(k')$ in the squeezed limit do not have explicit
dependences on $\alpha_{\rm s}$.

Substituting Eqs.~(\ref{eq:3-22a})--(\ref{eq:3-22d}) into
Eqs.~(\ref{eq:3-3b}) and (\ref{eq:3-21}), we have
\begin{align}
  \varDelta b^{\rm loc.}(k) &\approx
  \frac{ 2 f_{\rm NL}}{{\cal M}(k)}
  \int \frac{d^3k'}{(2\pi)^3} \tilde{c}^{\rm L}_2(k') P_{\rm L}(k'),
\label{eq:3-23a}\\
  \varDelta b^{\rm eql.}(k) &\approx
  \frac{6 f_{\rm NL} k^2}{{\cal M}(k)}
  \left[
      k^{2\alpha_{\rm s}} \int \frac{d^3k'}{(2\pi)^3}
      \frac{ \tilde{c}^{\rm L}_2(k')}{k'^{2(1+\alpha_{\rm s})}}
        P_{\rm L}(k')
  \right.
\nonumber\\
  &\hspace{3.9pc}
  \left.
      -\frac{(1+\alpha_{\rm s})^2}{3} \int \frac{d^3k'}{(2\pi)^3}
      \frac{ \tilde{c}^{\rm L}_2(k')}{k'^2} P_{\rm L}(k')
  \right]
\label{eq:3-23b}\\
  \varDelta b^{\rm fol.}(k) &\approx
    \frac{ 3 f_{\rm NL} k}{{\cal M}(k)}
  \int \frac{d^3k'}{(2\pi)^3} \frac{ \tilde{c}^{\rm L}_2(k')}{k'}
   P_{\rm L}(k'),
\label{eq:3-23c}\\
  \varDelta b^{\rm ort.}(k) &\approx
    -\frac{ 6 f_{\rm NL} k}{{\cal M}(k)}
  \int \frac{d^3k'}{(2\pi)^3} \frac{ \tilde{c}^{\rm L}_2(k')}{k'}
   P_{\rm L}(k'),
\label{eq:3-23d}
\end{align}
respectively. To derive the above equations, we use an approximation
${\cal M}(|\bm{k}-\bm{k}'|) \approx {\cal M}(k')$ in the squeezed
limit. In the case of equilateral non-Gaussianity, the asymptote of
Eq.~(\ref{eq:3-23b}) is somehow complicated for a general value of the
spectral index $n_{\rm s}$. In the case of scale-free spectrum,
$n_{\rm s}=1$ ($\alpha_{\rm s}=0$), Eq.~(\ref{eq:3-23b}) reduces to a
simpler form,
\begin{equation}
  \varDelta b^{\rm eql.}(k) \approx
    \frac{ 4 f_{\rm NL} k^2}{{\cal M}(k)}
  \int \frac{d^3k'}{(2\pi)^3}
   \frac{\tilde{c}^{\rm L}_2(k')}{k'^2}
    P_{\rm L}(k'), \quad (n_{\rm s} = 1).
\label{eq:3-24}
\end{equation}

The integrals in Eqs.~(\ref{eq:3-23a})--(\ref{eq:3-24}) do not depend
on $k$. Consequently, only the amplitudes of the scale-dependent bias
$\varDelta b(k)$ are sensitive to details of biasing in the
large-scale limit. The scaling indices are sensitive to only the
primordial bispectra, and are independent on details of biasing. On
sufficiently large scales, where $T(k) \approx 1$ and ${\cal M}(k)
\propto k^{-2}$, the known scalings
\begin{align}
  \varDelta b^{\rm loc.} &\propto k^{-2},
\label{eq:3-25a}\\
  \varDelta b^{\rm eql.} &\propto k^0,
\label{eq:3-25b}\\
  \varDelta b^{\rm fol.} &\propto k^{-1},
\label{eq:3-25c}\\
   \varDelta b^{\rm ort.} &\propto k^{-1}
\label{eq:3-25d}
\end{align}
hold irrespective to bias models. The scale-independence in the case
of equilateral model holds only for the scale-free power spectrum,
$n_{\rm s} = 1$. The above scalings are derived in the literature for
individual models of biasing, such as the halo model with
peak-background split \cite{SK10}, the high-peak model \cite{VM09},
and the local bias model \cite{TKM08}. Our general argument here shows
that those scalings are general consequences of squeezed limits of
primordial bispectra, and are independent on bias models. Only the
proportional coefficients of Eqs.~(\ref{eq:3-25a})--(\ref{eq:3-25d})
depend on bias models. This finding explains why different models of
bias in the literature give different amplitudes and the same spectral
index of scale-dependent bias.

For a given model of primordial bispectrum, obtaining the value of
scaling index of the scale-dependent bias in the large-scale limit is
straightforward even if the bias model is not specified: the scaling
index in that limit is just given by the squeezed limit of the
primordial bispectrum through Eq.~(\ref{eq:3-21}), i.e., the
large-scale behavior of $\varDelta b(k)$ as a function of $k$ is
determined by a combination $B_{\rm L}(k,k',|\bm{k}-\bm{k}'|)/P_{\rm
  L}(k)$ in the limit of $k \ll k'$. In the following sections, we
look into the amplitudes which are determined by concrete forms of the
renormalized bias function of second order, $c^{\rm
  L}_2(\bm{k}_1,\bm{k}_2)$.

\section{\label{sec:BiasFn}Shapes of renormalized bias functions}

Detailed amplitude and shape of scale-dependent bias depend on the
form of renormalized bias functions $c^{\rm L}_n$. In the lowest-order
approximation of this paper, the first two functions, $c^{\rm
  L}_1(\bm{k})$ and $c^{\rm L}_2(\bm{k}_1,\bm{k}_2)$, are of primary
interest. In this section, we derive the general form of renormalized
bias functions $c^{\rm L}_n$ in the halo model of bias.

\subsection{\label{subsec:HaloApp}
Asymptotes of the halo bias functions
}

The scale-dependent bias is previously derived in the halo approach
with the method of peak-background split
\cite{dal08,slo08,SK10,DJS11b}. With this method, it is shown that the
lowest-order term in scale-dependent bias, which is proportional to
$f_{\rm NL}$, is proportional to the first-order Lagrangian bias
parameter $b^{\rm L}_1$. On the other hand, Eulerian local bias models
predict the corresponding term is proportional to the second-order
bias parameter $b_2$ \cite{TKM08}. Both predictions agree with each
other in the high-peak limit, although the difference should be
important in physically realistic situations. Numerical simulations
indicate that the corresponding term is proportional to $b^{\rm L}_1$,
which agrees with the results of the peak-background split
\cite{DJS11a,DJS11b,SHMC11}.

In our asymptotic prediction of Eq.~(\ref{eq:3-21}), the corresponding
term is related to the second-order renormalized bias function,
$c^{\rm L}_2$. Does that mean our prediction contradicts with the
prediction of the peak-background split in the halo model? As we shall
see below, the answer is no. To see the relation between our general
results and the prediction of peak-background split, one needs to
derive the scale-dependence of the function $c^{\rm
  L}_2(\bm{k}_1,\bm{k}_2)$ in the model of halo bias.

First we introduce our notations of the halo approach. In the halo
model, the mass $M$ of halo is related to the Lagrangian radius $R$ of
a spherical cell by
\begin{equation}
  M = \frac43 \pi \bar{\rho}_0 R^3,
\label{eq:4-1}
\end{equation}
where $\bar{\rho}_0$ is the mean matter density at the present
time, or
\begin{equation}
  R = \left[
      \frac{M}{1.162\times 10^{12} h^{-1} M_\odot \varOmega_{\rm m0}}
      \right]^{1/3},
\label{eq:4-2}
\end{equation}
where $M_\odot =1.989\times 10^{30}\,{\rm kg}$ is the mass of the sun.
In the following, the radius $R$ is always a function of a mass scale
$M$ through the above equation. The density variance of mass scale $M$
is given by
\begin{equation}
  {\sigma_M}^2 = \int \frac{d^3k}{(2\pi)^3} W^2(kR) P_{\rm L}(k),
\label{eq:4-3}
\end{equation}
where the function $W(kR)$ is usually chosen to be a top-hat window
function,
\begin{equation}
    W(x) = \frac{3j_1(x)}{x} = \frac{3\sin x - 3x \cos x}{x^3},
\label{eq:4-3-1}
\end{equation}
and $j_1(x)$ is the first-order spherical Bessel function.

In Ref.~\cite{mat11}, the large-scale asymptotes of the bias functions
are derived in the halo approach with a universal mass function. 
The mass function is said to be universal when the mass function has a
form
\begin{equation}
  n(M)dM = \frac{\bar{\rho}_0}{M} f_{\rm MF}(\nu) \frac{d\nu}{\nu},
\label{eq:4-4}
\end{equation}
where $\nu = \delta_{\rm c}/\sigma_M$ is a function of mass $M$, and
$\delta_{\rm c}$ is the critical overdensity for spherical collapse.
In the Einstein-de~Sitter model, the critical overdensity is exactly
independent of redshift, $\delta_{\rm c} = 3(3\pi/2)^{2/3}/5 \simeq
1.686$, while it only weakly depends on cosmological parameters and
redshift in general cosmology. The multiplicity function $f_{\rm
  MF}(\nu)$ has a normalization
\begin{equation}
  \int_0^\infty f_{\rm MF}(\nu)\, \frac{d\nu}{\nu} = 1
\label{eq:4-5}
\end{equation}
to ensure all the mass in the universe is to be contained in halos.

In the literature, several forms of the multiplicity function $f_{\rm MF}$
are proposed. In the original Press-Schechter (PS) theory \cite{PS74},
it has a form,
\begin{equation}
  f_{\rm PS}(\nu) = \sqrt{\frac{2}{\pi}}\,\nu e^{-\nu^2/2}.
\label{eq:4-6}
\end{equation}
Sheth and Tormen (ST) \cite{ST99} give a better fit to numerical
simulations of cold-dark-matter type cosmologies with Gaussian initial
conditions,
\begin{equation}
  f_{\rm ST}(\nu) = A(p) \sqrt{\frac{2}{\pi}}
  \left[1 + \frac{1}{(q\nu^2)^p}\right]\sqrt{q}\,\nu e^{-q\nu^2/2},
\label{eq:4-7}
\end{equation}
where $p=0.3$, $q=0.707$ are numerically fitted parameters, and $A(p)
= [1+\pi^{-1/2}2^{-p}\varGamma(1/2-p)]^{-1}$ is the normalization
factor. When $p=0$, $q=1$, the ST mass function reduces to the PS mass
function.

Several other fitting formulas for $f_{\rm MF}$ have been proposed
with numerically improved calibrations \cite{Jen01,War06,Cro10}. In
Warren et al.~\cite{War06}, for example, the multiplicity function is
fitted as a function of $\sigma \equiv \sigma_M$ instead of $\nu$ as
\begin{equation}
  \tilde{f}_{\rm W}(\sigma) = A \left(\sigma^{-a} + b\right)
          \exp\left(-\frac{c}{\sigma^2}\right),
\label{eq:4-8}
\end{equation}
where $A,a,b,c$ are fitting parameters. The same functional form is
applied to MICE simulations in Ref.~\cite{Cro10}, allowing the
parameters redshift-dependent. Their values are given by $A(z) =
0.58(1+z)^{-0.13}$, $a(z) = 1.37(1+z)^{-0.15}$, $b(z) =
0.3(1+z)^{-0.084}$, $c(z)=1.036(1+z)^{-0.024}$. As a function of
$\nu$, Eq.~(\ref{eq:4-8}) can be re-expressed as
\begin{equation}
  f_{\rm W}(\nu) = \tilde{f}_{\rm W}(\delta_{\rm c}/\nu)
  = A \left[\left(\frac{\nu}{\delta_{\rm c}}\right)^a + b\right]
          \exp\left(-\frac{c\nu^2}{{\delta_{\rm c}}^2}\right).
\label{eq:4-9}
\end{equation}
In the following, we refer to the above form as ``MICE mass function''
when the redshift-dependent parameters with MICE simulations are
adopted. When the parameters $A,a,b,c$ are redshift-dependent as in
the case of MICE simulations, the multiplicity function explicitly
depends on the redshift as $f_{\rm MF}(\nu,z)$. In this case, the mass
function is not universal anymore. Throughout this paper, explicit
dependences on time is notationally suppressed in the arguments of
functions, and we adopt the notation $f_{\rm MF}(\nu)$ even if this
function explicitly depends on the redshift. The PS mass function is
recovered when we formally substitute $A = \sqrt{2/\pi}\,\delta_{\rm
  c}$, $a=1$, $b=0$, $c={\delta_{\rm c}}^2/2$ into Eq.~(\ref{eq:4-9}).

Using the notations introduced above, the long-wavelength asymptotes
of the bias functions derived in Ref.~\cite{mat11} have the form,
\begin{equation}
  c^{\rm L}_n(\bm{k}_1,\ldots,\bm{k}_n) \approx b^{\rm L}_n(M)
 \quad (|\bm{k}_i| \rightarrow 0),
\label{eq:4-10}
\end{equation}
where $b^{\rm L}_n(M)$ is a scale-independent function defined by
\begin{equation}
  b^{\rm L}_n(M) =
  \left(-\frac{1}{\sigma_M}\right)^n
      \frac{f_{\rm MF}^{(n)}(\nu)}{f_{\rm MF}(\nu)},
\label{eq:4-11}
\end{equation}
and $f_{\rm MF}^{(n)} = d^nf_{\rm MF}/d\nu^n$ denotes $n$th derivative
with respect to $\nu$.

\begin{figure}
\begin{center}
\includegraphics[width=20.5pc]{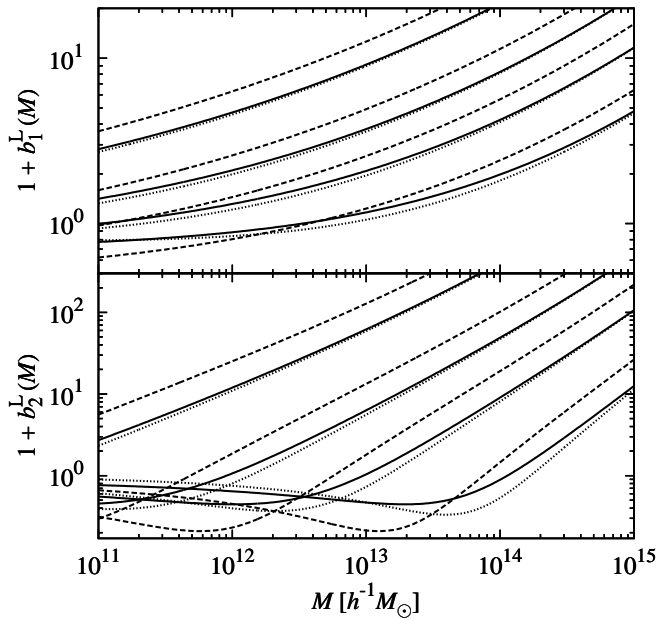}
\caption{\label{fig:Bias} The scale-independent Lagrangian bias
  parameters derived from the Sheth-Tormen (solid lines),
  Press-Schechter (dashed lines) and MICE (dotted lines) mass
  functions for $z=0,1,2,4$ (from bottom to top on large mass scales).
  The unity is added to each parameter to show the negative values in
  this logarithmic plot.
}
\end{center}
\end{figure}

Specifically, when the PS mass function with Eq.~(\ref{eq:4-6}) is
applied, we have
\begin{equation}
  b^{\rm L}_1(M) = \frac{\nu^2 - 1}{\delta_{\rm c}}, \quad
  b^{\rm L}_2(M) = \frac{\nu^4 - 3\nu^2}{{\delta_{\rm c}}^2},
\label{eq:4-12}
\end{equation}
and so forth, which are consistent with the results derived from the
model of spherical collapse \cite{MW96,MJW97}. When the ST mass
function with Eq.~(\ref{eq:4-7}) is applied, we have
\begin{align}
  b^{\rm L}_1(M)
  &= \frac{1}{\delta_{\rm c}}
  \left[q\nu^2 - 1 + \frac{2p}{1 + (q\nu^2)^p}\right],
\label{eq:4-13a}\\
  b^{\rm L}_2(M)
  &= \frac{1}{{\delta_{\rm c}}^2}
  \left[q^2\nu^4 - 3 q\nu^2 + \frac{2p(2q\nu^2 + 2p -1)}{1 +
        (q\nu^2)^p}\right],
\label{eq:4-13b}
\end{align}
and so forth. When the MICE mass function with Eq.~(\ref{eq:4-8}) is
applied, we have
\begin{align}
  b^{\rm L}_1(M) &= \frac{1}{\delta_{\rm c}}
  \left( \frac{2c}{\sigma^2} - \frac{a}{1+b\sigma^a}\right),
\label{eq:4-14a}\\
  b^{\rm L}_2(M) &= \frac{1}{{\delta_{\rm c}}^2}
  \left[ \frac{4c^2}{\sigma^4} - \frac{2c}{\sigma^2}
      - \frac{a\left(4c/\sigma^2 - a + 1\right)}{1+b\sigma^a}
  \right],
\label{eq:4-14b}
\end{align}
and so forth, where $\sigma = \sigma_M = \delta_{\rm c}/\nu$. The bias
parameters $b^{\rm L}_1(M)$ and $b^{\rm L}_2(M)$ as functions of mass
are plotted in Fig.~\ref{fig:Bias} for three different mass functions
considered above.

If the asymptotes of Eq.~(\ref{eq:4-10}) are naively used in
Eq.~(\ref{eq:3-21}), the scale-dependent bias is proportional to the
second-order Lagrangian bias parameter $\varDelta b \propto b^{\rm
  L}_2$. Bias parameters are scale-independent in models of local
bias. Therefore, the use of asymptotes of Eq.~(\ref{eq:4-10}) restrict
ourselves to a model of local bias. Local bias models predict that the
scale-dependent bias $\varDelta b$ is proportional to the second-order
bias parameter $b_2$ in general \cite{TKM08}. One should note that
when $c^{\rm L}_2$ is exactly constant, the integral in
Eq.~(\ref{eq:3-23a}) logarithmically diverges for the local model of
primordial non-Gaussianity with a cold-dark-matter type power spectrum
which has a small-scale asymptote $P_{\rm L}(k) \propto k^{-3}$ for
$k\rightarrow\infty$.

The local bias model turned out {\em not} to be a good approximation
in modeling the scale-dependent bias of halos measured in numerical
simulations with non-Gaussian initial conditions
\cite{GP10,DJS11a,SHMC11}. The scale-dependent bias is more or less
proportional to the first-order bias parameter $\varDelta b \propto
b^{\rm L}_1$ rather than the second-order parameter $b^{\rm L}_2$. The
property $\varDelta b \propto b^{\rm L}_1$ is a general prediction of
the halo approach with the method of peak-background split
\cite{dal08,slo08,SK10}. That means the method of peak-background
split takes into account the nonlocal nature of biasing. The fact that
the predictions of the peak-background split are better than those of
local bias model in reproducing the results of numerical simulations
implies that nonlocal nature of biasing is important to understand the
scale-dependent bias from the primordial non-Gaussianity.

\subsection{\label{subsec:HaloBias}
Renormalized bias functions of halos without assuming peak-background split
}

In the following several subsections, we extend the calculation of the
renormalized bias functions beyond the asymptotic limit of
Eq.~(\ref{eq:4-10}), without resorting to the approximation of
peak-background split. In the end, the known formula of
scale-dependent bias derived from the peak-background split is exactly
re-derived as limiting cases, which is one of the remarkable findings
in this paper.

In the literature, analytic formulas of halo bias are derived by more or
less adopting a concept of peak-background split. This method is
necessary because the halo approach is based on a statistical nature
of extended Press-Schechter mass function. In such an approach, the
local mass function is obtained by averaging over small-scale
fluctuations, while large-scale fluctuations are considered as
background modulation field, which leads spatial fluctuations of
number density of halos. Comparing the fluctuations of the halo number
density field and those of mass, the halo bias is analytically
derived.

However, the biasing can be seen as a deterministic process at a most
fundamental level, in which any statistical information is not
required. One can think of getting a halo catalog in numerical
simulations to understand the situation. Just one realization of the
initial condition deterministically gives subsequent nonlinear
evolutions and formation sites of halos.

When only leading growing modes are considered in a perturbation
theory, any structure in the universe is deterministically related to
the linear density field. The biasing relation should not require
statistical information of the field. In calculating the renormalized
bias functions of Eq.~(\ref{eq:2-3}), the number density fluctuations
$\delta^{\rm L}_{\rm X}$ is a deterministic functional of linear
density field $\delta_{\rm L}$. Any statistical quantities, such as
the short-mode power spectrum in the method of peak-background split,
are not expected to appear at the most fundamental level.

From considerations above, we start from an unaveraged version of the
original Press-Schechter (PS) formalism \cite{PS74}. The linear
density field smoothed over mass scale $M$,
\begin{equation}
  \delta_M(\bm{x})
  = \int \frac{d^3k}{(2\pi)^3}
  e^{i\bm{k}\cdot\bm{x}}W(kR) \delta_{\rm L}(\bm{k}),
\label{eq:4-21}
\end{equation}
is a fundamental element in the original PS formalism and its
variants. When $P(M,\delta_{\rm c})$ denotes the probability that (or the
volume fraction where) the value $\delta_M$ exceeds a critical value
$\delta_{\rm c}$, the averaged number density of halos in the original
PS formalism are given by
\begin{equation}
  n(M) = -\frac{2\bar{\rho}_0}{M}
  \frac{\partial}{\partial M}P(M,\delta_{\rm c}),
\label{eq:4-22}
\end{equation}
where $n(M)$ is the {\em differential} mass function. In some
literatures, $n(M)$ is denoted by $dn/dM$, which notation we do not
adopt. When the distribution function of $\delta_M$ is exactly
Gaussian, Eq.~(\ref{eq:4-22}) is equivalent to Eq.~(\ref{eq:4-4})
with Eq.~(\ref{eq:4-6}).

For our purpose, we need to have a local number density of halos
$n(\bm{x},M)$, instead of spatially averaged one. The condition that a
mass element at a particular point is included in a collapsed halo of
mass greater than $M$ is not stochastic but deterministic when the
linear density field is given. Consequently, localized version of
Eq.~(\ref{eq:4-22}) should be
\begin{equation}
  n(\bm{x},M) = -\frac{2\bar{\rho}_0}{M}
  \frac{\partial}{\partial M}\varTheta[\delta_M(\bm{x}) - \delta_{\rm c}],
\label{eq:4-23}
\end{equation}
where $\varTheta(x)$ is a step function, and $\bm{x}$ is a Lagrangian
position. The statistical quantities do not appear in this relation.
For a given realization of linear density field $\delta_{\rm L}$, the
mass element at a given point is either collapsed $\delta_M >
\delta_{\rm c}$ or uncollapsed $\delta_M < \delta_{\rm c}$. Taking the
spatial average of Eq.~(\ref{eq:4-23}), Eq.~(\ref{eq:4-22}) of the
original PS formalism follows, because
\begin{equation}
  \langle\varTheta(\delta_M(\bm{x})-\delta_{\rm c})\rangle
  = P(M,\delta_{\rm c}).
\label{eq:4-24}
\end{equation}
In a picture of excursion set approach \cite{BCEK91}, the step
function in Eq.~(\ref{eq:4-23}) should be replaced by an operator
$N_{\rm 1up}(M)$, which is zero until the first up-crossing $\delta_M >
\delta_{\rm c}$ occurs as the mass scale $M$ decreases from infinity,
and it becomes unity below that mass scale. This operator $N_{\rm
  1up}$ is not just a single function of $\delta_M$, and analytic
treatments are more complicated if not impossible. In this paper, we
just use Eq.~(\ref{eq:4-23}) in the following consideration for
simplicity.

It is also possible to consider the function $\varTheta$ is a general
function which is not necessarily a step function. When this function
is a step function, the PS mass function exactly follows when the
linear density field is random Gaussian and the model of spherical
collapse is literally assumed. However, in reality, the mass function
is different from the PS one even if the linear density field is
random Gaussian. In approaches of universal mass function, the step
function is deformed to a different function to reproduce a mass
function $n(M)$ in numerical simulations. We allow this possibility
and the function $\varTheta$ is not necessarily a step function in the
following derivation.

The number density contrast in Lagrangian space is given by
$\delta^{\rm L}_{\rm h}(\bm{x}) = n(\bm{x},M)/n(M) - 1$ for halos of
mass $M$, where Eqs.~(\ref{eq:4-22}) and (\ref{eq:4-23}) are assumed.
Taking functional derivatives of Eq.~(\ref{eq:4-23}), we have
\begin{multline}
  \frac{\delta^n n(\bm{x},M)}
  {\delta\delta_{\rm L}(\bm{k}_1)\cdots\delta\delta_{\rm L}(\bm{k}_n)}
  = -\frac{e^{i(\bm{k}_1 + \cdots + \bm{k}_n)\cdot\bm{x}}}
  {(2\pi)^{3n}}\frac{2\bar{\rho}_0}{M}
\\ \times
    \frac{\partial}{\partial M}
    \left[\varTheta^{(n)}(\delta_M-\delta_{\rm c})
    W(k_1R) \cdots W(k_nR)\right],
\label{eq:4-25}
\end{multline}
where Eq.~(\ref{eq:4-21}) is used and $\varTheta^{(n)}(x) =
d^n\varTheta(x)/dx^n$. Fourier transforming the above equation with
respect to $\bm{x}$ and taking ensemble average, the renormalized bias
functions of Eq.~(\ref{eq:2-3}) or Eq.~(\ref{eq:2-4}) reduce to
\begin{equation}
  c^{\rm L}_n(\bm{k}_1,\ldots,\bm{k}_n) =
  \frac{\displaystyle
    (-1)^n
    \frac{\partial}{\partial M}
    \left[
    \frac{\partial^n P(M,\delta_{\rm c})}{{\partial\delta_{\rm c}}^n}
      W(k_1R) \cdots W(k_nR)\right]}
  {\displaystyle
    \frac{\partial P(M,\delta_{\rm c})}{\partial M}},
\label{eq:4-26}
\end{equation}
where $R$ is an explicit function of $M$ and partial derivatives of
the window functions with respect to $M$ do not vanish.

\subsection{\label{subsec:UMF}
Renormalized bias functions in universal mass functions
}

Assuming a universal mass function of Eq.~(\ref{eq:4-4}), we have a
form
\begin{equation}
  P(M,\delta_{\rm c}) = \frac12 F(\nu),
\label{eq:4-27a}
\end{equation}
where
\begin{equation}
  F(\nu) \equiv \int_\nu^\infty\frac{f_{\rm MF}(\nu)}{\nu} d\nu,
\label{eq:4-27b}
\end{equation}
is a filling factor of collapsed regions. Partial derivatives in
Eq.~(\ref{eq:4-26}) are given by
\begin{align}
  \frac{\partial P(M,\delta_c)}{\partial M}
  &= \frac{f_{\rm MF}(\nu)}{2} \frac{d\ln\sigma_M}{dM},
\label{eq:4-28a}\\
  \frac{\partial^n P(M,\delta_{\rm c})}{{\partial\delta_{\rm c}}^n} &= 
  \frac{F^{(n)}(\nu)}{2{\sigma_M}^n},
\label{eq:4-28b}
\end{align}
where
\begin{align}
  F^{(n)}(\nu) &= \frac{d^n F}{d\nu^n}
  = - \frac{d^{n-1}}{d\nu^{n-1}}
  \left[\frac{f_{\rm MF}(\nu)}{\nu}\right]
\nonumber\\
  &= \frac{(-1)^n (n-1)!}{\nu^n} \sum_{j=0}^{n-1} \frac{(-1)^j}{j!}
  \nu^j f_{\rm MF}^{(j)}(\nu).
\label{eq:4-29}
\end{align}
In this case, Eq.~(\ref{eq:4-26}) reduces to
\begin{equation}
  c^{\rm L}_n(\bm{k}_1,\ldots,\bm{k}_n) =
  \frac{(-1)^n}{f_{\rm MF}(\nu)}
  \frac{d}{d\ln\sigma_M}
  \left[
    \frac{F^{(n)}(\nu)W(k_1R) \cdots W(k_nR)}{{\sigma_M}^n}
  \right].
\label{eq:4-30}
\end{equation}

To evaluate the above expression, we have useful formulas,
\begin{align}
&  \frac{dF^{(n)}(\nu)}{d\ln\sigma_M} = -\nu F^{(n+1)}
  = \frac{(-1)^n n!}{\nu^n} \sum_{j=0}^{n} \frac{(-1)^j}{j!}
  \nu^j f_{\rm MF}^{(j)},
\label{eq:4-31a}\\
&  \frac{d}{d\ln\sigma_M}\left(\frac{F^{(n)}(\nu)}{{\sigma_M}^n}\right)
  = \frac{1}{{\sigma_M}^n}
  \left[\frac{dF^{(n)}(\nu)}{d\ln\sigma_M} - nF^{(n)}(\nu)\right]
  = \frac{f_{\rm MF}^{(n)}}{{\sigma_M}^n}.
\label{eq:4-31b}
\end{align}

\begin{figure}
\begin{center}
\includegraphics[width=20.5pc]{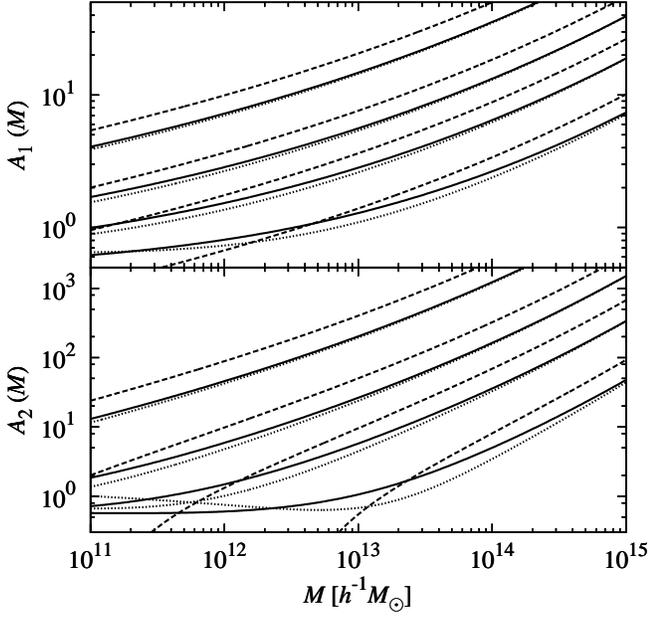}
\caption{\label{fig:aonetwo} The functions $A_1(M)$ and $A_2(M)$
  derived from the Sheth-Tormen (solid lines), Press-Schechter (dashed
  lines) and MICE (dotted lines) mass functions for $z=0,1,2,4$ (from
  bottom to top on large mass scales).
}
\end{center}
\end{figure}

One can use the scale-independent parameters $b^{\rm L}_n(M) =
(-1/\sigma_M)^nf_{\rm MF}^{(n)}/f_{\rm MF}$ introduced in
Eq.~(\ref{eq:4-11}) in the above equations. Accordingly,
Eq.~(\ref{eq:4-30}) reduces to two equivalent expressions,
\begin{multline}
  c^{\rm L}_n(\bm{k}_1,\ldots,\bm{k}_n)
  = \frac{A_n(M)}{{\delta_{\rm c}}^n} W(k_1R) \cdots W(k_nR)
\\
  + \frac{A_{n-1}(M)\, {\sigma_M}^n}{{\delta_{\rm c}}^n}
  \frac{d}{d\ln\sigma_M}
    \left[\frac{W(k_1R) \cdots W(k_nR)}{{\sigma_M}^n} \right],
\label{eq:4-32}
\end{multline}
and
\begin{multline}
  c^{\rm L}_n(\bm{k}_1,\ldots,\bm{k}_n) =
  b^{\rm L}_n(M) W(k_1R) \cdots W(k_nR)
\\
  + \frac{A_{n-1}(M)}{{\delta_{\rm c}}^n}
  \frac{d}{d\ln\sigma_M}
    \left[W(k_1R) \cdots W(k_nR)\right].
\label{eq:4-33}
\end{multline}
where
\begin{equation}
  A_n(M) \equiv
  \sum_{j=0}^n \frac{n!}{j!}{\delta_{\rm c}}^j\, b^{\rm L}_j(M).
\label{eq:4-34}
\end{equation}
and $b^{\rm L}_0(M) \equiv 1$ for consistency.
There is a recursion relation,
\begin{equation}
  A_n = nA_{n-1} + {\delta_{\rm c}}^n b^{\rm L}_n,
\label{eq:4-35}
\end{equation}
which also guarantees the equivalence between the two expressions of
Eqs.~(\ref{eq:4-32}) and (\ref{eq:4-33}).
In this paper, we need the renormalized bias functions up to second
order. Explicitly, we have
\begin{align}
  A_0(M) &= 1,
\label{eq:4-36a}\\
  A_1(M) &= 1 + \delta_{\rm c} b^{\rm L}_1(M),
\label{eq:4-36b}\\
  A_2(M) &= 2  + 2 \delta_{\rm c} b^{\rm L}_1(M)
  + {\delta_{\rm c}}^2 b^{\rm L}_2(M).
\label{eq:4-36c}
\end{align}
The functions $A_1(M)$ and $A_2(M)$ are plotted in
Fig.~\ref{fig:aonetwo} for three kinds of mass functions considered in
Sec.~\ref{subsec:HaloApp}.

When all the magnitudes of wavevector are small,
$|\bm{k}_i|\rightarrow 0$, the window function becomes asymptotically
unity, $W(k_iR)\rightarrow 1$. Accordingly, the previous result of
Eq.~(\ref{eq:4-10}) is correctly recovered in this limit. Therefore,
Eq.~(\ref{eq:4-32}) or Eq.~(\ref{eq:4-33}) gives the form of
renormalized bias functions beyond the approximation of large-scale
limit. Scale-dependence of bias functions means nonlocality of the
bias in general. The derived renormalized bias functions are
scale-dependent on scales of mass $M$, which suggests that the halo
bias is nonlocal on scales of halo mass, as naturally expected. Our
formula, Eq.~(\ref{eq:4-32}) or (\ref{eq:4-33}), can be used in
general applications of iPT with the halo bias, beyond the problem of
scale-dependent bias in this paper.

Up to the second order, we explicitly have
\begin{align}
  c^{\rm L}_1(k) &= b^{\rm L}_1(M) W(kR)
  + \frac{1}{\delta_{\rm c}} \frac{\partial W(kR)}{\partial\ln\sigma_M},
\label{eq:4-37a}\\
  c^{\rm L}_2(\bm{k}_1,\bm{k}_2) &=
  b^{\rm L}_2(M) W(k_1R) W(k_2R)
\nonumber\\
  & \quad + \frac{1 + \delta_{\rm c} b^{\rm L}_1(M)}{{\delta_{\rm c}}^2}
  \frac{\partial}{\partial\ln\sigma_M}\left[W(k_1R)W(k_2R)\right].
\label{eq:4-37b}
\end{align}

The Eqs.~(\ref{eq:4-37a}) and (\ref{eq:4-37b}) complete the elements
to calculate our prediction for the scale-dependent bias,
Eq.~(\ref{eq:2-13}), in the case of halo bias. For convenience, we define
\begin{equation}
  {\cal I}_n(k) = \int \frac{d^3k'}{(2\pi)^3}
  \hat{\cal I}_n(\bm{k},\bm{k}') B_{\rm L}(k,k',|\bm{k}-\bm{k}'|),
\label{eq:4-38}
\end{equation}
for $n=1,2$, where
\begin{align}
  \hat{\cal I}_1(\bm{k},\bm{k}') &=
  \frac{2\bm{k}\cdot\bm{k}'}{k'^2} W(|\bm{k}-\bm{k}'|R),
\label{eq:4-39a}\\
  \hat{\cal I}_2(\bm{k},\bm{k}') &= W(k'R)W(|\bm{k}-\bm{k}'|R).
\label{eq:4-39b}
\end{align}
When the bias functions are given by Eqs.~(\ref{eq:4-37a}) and
(\ref{eq:4-37b}), the functions $Q_1(k)$ and $Q_2(k)$ of
Eqs.~(\ref{eq:2-8}), (\ref{eq:2-9b}), (\ref{eq:2-9c}) are represented as
\begin{align}
  Q_1(k) &= b^{\rm L}_1 {\cal I}_1(k) + \frac{1}{\delta_{\rm c}}
  \frac{\partial{\cal I}_1(k)}{\partial\ln\sigma_M},
\label{eq:4-39-1a}\\
  Q_2(k) &= b^{\rm L}_2 {\cal I}_2(k)
  + \frac{1+\delta_{\rm c}b^{\rm L}_1}{{\delta_{\rm c}}^2}
  \frac{\partial{\cal I}_2(k)}{\partial\ln\sigma_M}.
\label{eq:4-39-1b}
\end{align}

\subsection{\label{subsec:MassRange}
Effects of Mass Selection
}

The above calculations assume that all halos have the same mass, $M$.
In reality, halos of different masses are contained in a given sample.
When the finiteness of mass range is not negligible, the global and
local number densities of halos are given by
\begin{align}
  N(M) &= \int dM\phi(M)n(M),
\label{eq:4-101a}\\
  N(\bm{x},M) &= \int dM\phi(M)n(\bm{x},M),
\label{eq:4-101b}
\end{align}
respectively, where $\phi(M)$ is an arbitrary selection function of
mass. When the mass of halos in a range $M_1 < M < M_2$ are evenly
selected, the selection function is given by $\phi(M) = 1$ when $M_1 <
M < M_2$ and $\phi(M) = 0$ otherwise. When the mass range is
negligibly small, $\phi(M') = \delta_{\rm D}(M'-M)$.

The number density contrast in Lagrangian space is given by
$\delta^{\rm L}_{\rm h}(\bm{x}) = N(\bm{x},M)/N(M)-1$. Following a
similar procedure to obtain Eq.~(\ref{eq:4-26}), we have
\begin{multline}
  c^{\rm L}_n(\bm{k}_1,\ldots,\bm{k}_n)
\\ =
  \frac{\displaystyle
    (-1)^n \int dM \frac{\phi(M)}{M}
    \frac{\partial}{\partial M}
    \left[
    \frac{\partial^n P(M,\delta_{\rm c})}{{\partial\delta_{\rm c}}^n}
      W(k_1R) \cdots W(k_nR)\right]}
  {\displaystyle  \int dM \frac{\phi(M)}{M}
    \frac{\partial P(M,\delta_{\rm c})}{\partial M}}.
\label{eq:4-102}
\end{multline}
instead of Eq.~(\ref{eq:4-26}). Taking account of Eqs.~(\ref{eq:4-22})
and (\ref{eq:4-26}), we find
\begin{equation}
  c^{\rm L}_n(\bm{k}_1,\ldots,\bm{k}_n) =
  \frac{\displaystyle
    \int dM\phi(M)n(M) c^{\rm L}_n(\bm{k}_1,\ldots,\bm{k}_n;M)}
  {\displaystyle
    \int dM\phi(M)n(M)},
\label{eq:4-103}
\end{equation}
where $c^{\rm L}_n(\bm{k}_1,\ldots,\bm{k}_n;M)$ is the renormalized
bias function for a fixed mass $M$ derived in the previous subsection,
Eqs.~(\ref{eq:4-30}), (\ref{eq:4-32}) or (\ref{eq:4-33}). Quite
naturally, Eq.~(\ref{eq:4-103}) shows that the renormalized bias
functions for mass-selected halo samples are obtained by averaging
over mass-dependent bias functions weighted by selected number of
halos.

\subsection{\label{subsec:PScancel} Cancellation of highest-order bias
  parameters in Press-Schechter mass function}

In the case of PS mass function, there is a remarkable property that
the $n$th-order renormalized bias function $c^{\rm L}_n$ only depend
on next two lower-order bias parameters $b^{\rm L}_{n-1}$ and $b^{\rm
  L}_{n-2}$, as shown below.

The scale-independent parameter $b^{\rm L}_n$ of Eq.~(\ref{eq:4-11})
in the case of PS mass function is given by
\begin{equation}
  b^{\rm L}_n(M) = \frac{\nu^{n-1}H_{n+1}(\nu)}{{\delta_{\rm c}}^n},
\label{eq:4-40}
\end{equation}
where $H_n(\nu) = e^{\nu^2/2}(-d/d\nu)^n e^{-\nu^2/2}$ denotes Hermite
polynomials. Using a recursion relation of the Hermite polynomials,
$H_{n+1}(\nu) = \nu H_n(\nu) - n H_{n-1}(\nu)$, we observe that
remarkable cancellations in the series of Eq.~(\ref{eq:4-34}) occur
and just a single term survives as
\begin{equation}
  A_n(M) = \nu^n H_n(\nu)
  = \frac{{\delta_{\rm c}}^{n+1}}{{\sigma_M}^2} b^{\rm L}_{n-1}(M),
\label{eq:4-41}
\end{equation}
where $b^{\rm L}_{-1}(M) \equiv {\sigma_M}^2/\delta_{\rm c}$ for
consistency. For $n=1,2$, we have
\begin{equation}
  A_1(M) = \nu^2,
  \quad
  A_2(M) = \nu^2 \delta_{\rm c} b^{\rm L}_1(M).
\label{eq:4-41-1}
\end{equation}

Interestingly, the highest-order parameter $b^{\rm L}_n$ vanishes in
$A_n$, and it is simply proportional to the next lower-order parameter
$b^{\rm L}_{n-1}$ in this case of PS mass function. Accordingly, the
renormalized bias function $c^{\rm L}_n$ of $n$th order depends only
on lower-order bias parameters $b^{\rm L}_{n-1}$ and $b^{\rm L}_{n-2}$
in the form of Eq.~(\ref{eq:4-32}) with Eq.~(\ref{eq:4-41}). In the
next subsection, this notable property turns out to be the reason why
the scale-dependent bias from the primordial bispectrum is roughly
proportional to the first-order bias parameter $b^{\rm L}_1$ instead of
the second-order parameter $b^{\rm L}_2$.

However, the exact cancellations do not occur in other cases than the
PS mass function. For example, in the case of ST mass function with
the bias parameters of Eqs.~(\ref{eq:4-13a}) and (\ref{eq:4-13b}),
coefficients of the renormalized bias functions, Eqs.~(\ref{eq:4-36b})
and (\ref{eq:4-36c}) reduce to
\begin{align}
  A_1(M) &= q\nu^2 + \frac{2p}{1 + (q\nu^2)^p},
\label{eq:4-42a}\\
  A_2(M) &= q \nu^2 \delta_{\rm c} b^{\rm L}_1(M)
  + \frac{2p(q\nu^2 + 2p + 1)}{1 + (q\nu^2)^p},
\label{eq:4-42b}
\end{align}
and so forth. The leading term of $A_n$ is still proportional to the
next lower-order bias parameter $b^{\rm L}_{n-1}$, but correction
terms, which are proportional to $p$, additionally appear. In the case
of PS mass function, $p=0$ and $q=1$, the above equations agree with
Eq.~(\ref{eq:4-41}) of $n=1,2$ as they should. Similarly, in the case
of MICE mass function with the bias parameters of
Eqs.~(\ref{eq:4-14a}) and (\ref{eq:4-14b}), we have
\begin{align}
   A_1(M) &= \frac{2c}{\sigma^2} + 1 - \frac{a}{1+b\sigma^a},
\label{eq:4-43a}\\
   A_2(M) &= \frac{2c}{\sigma^2} \delta_{\rm c} b^{\rm L}_1(M)
   + \frac{2c}{\sigma^2} + 2
   - \frac{a\left(2c/\sigma^2 - a + 3\right)}{1+b\sigma^a},
\label{eq:4-43b}
\end{align}
and so forth. Again, when we substitute $a=1$, $b=0$ and
$c={\delta_{\rm c}}^2/2$, the above equations agree with
Eq.~(\ref{eq:4-41}) of $n=1,2$.

\subsection{\label{subsec:RederivePBS}
Relation to the previous formula derived by the peak-background split
}

We explicitly show below that predictions of the peak-background split
are re-derived as a special case of the general formula we have
derived. First we drop the terms $Q_0$, $Q_1$ in Eq.~(\ref{eq:2-13}),
which are subdominant on large scales, and only consider the dominant
term $Q_2(k)$. In this approximation, the scale-dependent bias is
approximately given by $\varDelta b(k) \approx Q_2(k)/2P_{\rm L}(k)$,
i.e.,
\begin{equation}
  \varDelta b(k) \approx
  \frac{1}{2P_{\rm L}(k)}
  \int\frac{d^3k'}{(2\pi)^3} c^{\rm L}_2(\bm{k}',\bm{k}-\bm{k}')
  B_{\rm L}(k,k',|\bm{k}-\bm{k'}|).
\label{eq:4-50}
\end{equation}
Substituting the second-order renormalized bias function $c^{\rm
  L}_2$ in the form of Eq.~(\ref{eq:4-32}) with $n=2$ into the above
equation, we have
\begin{equation}
  \varDelta b(k) \approx 
  \frac{{\sigma_M}^2}{2{\delta_{\rm c}}^2}
  \left[
    A_2 {\cal I}(k)
    + A_1
    \frac{\partial{\cal I}(k)}{\partial\ln\sigma_M}
   \right],
\label{eq:4-51}
\end{equation}
where
\begin{align}
  {\cal I}(k) &\equiv \frac{{\cal I}_2(k)}{{\sigma_M}^2P_{\rm L}(k)} 
\nonumber\\
  &\approx \frac{1}{{\sigma_M}^2P_{\rm L}(k)}
  \int\frac{d^3k'}{(2\pi)^3} W^2(k'R) B_{\rm L}(k,k',|\bm{k}-\bm{k'}|).
\label{eq:4-52}
\end{align}

When bias parameters of the PS mass function are used, the
parameter $A_1$, $A_2$ are given by Eq.~(\ref{eq:4-41-1}), and
Eq.~(\ref{eq:4-51}) reduces to
\begin{equation}
  \varDelta b(k) \approx
   \frac12 \delta_{\rm c} b^{\rm L}_1 {\cal I}(k) + 
   \frac12 \frac{\partial{\cal I}(k)}{\partial\ln\sigma_M}.
\label{eq:4-53}
\end{equation}
This equation is exactly the same as a recent prediction of the
peak-background split \cite{DJS11a,DJS11b}. In the literature, the
window function $W(kR)$ is sometimes additionally divided into the
function ${\cal I}(k)$ because the bias is defined with respect to the
smoothed density field in the latter case (private communications with
V.~Desjacques). Earlier predictions in the literature
\cite{dal08,DS10a,SK10} are reproduced from the first term of
Eq.~(\ref{eq:4-53}). Although the second term is subdominant in the
high-peak limit, it is significant for most relevant peak heights
\cite{DJS11a,DJS11b}. In the local model of non-Gaussianity, the
second term vanishes on sufficiently large scales.

Importantly, our derivation does not use the approximation of the
peak-background split to obtain the same prediction. Therefore, the
known result of Eq.~(\ref{eq:4-53}) turns out not to particularly
depend on the approximation of the peak-background split. However, we
have found that this equation is consistent only with the PS mass
function. That is not surprising, because the arguments of
peak-background split use an auxiliary Gaussian field as an
fundamental field in the analysis. The original PS mass function is
properly derived by assuming the Gaussian statistics for the linear
density field. Applying the Gaussian statistics to the auxiliary field
is only consistent with the original PS mass function in the
peak-background split.

Our derivation, on the other hand, does not use any auxiliary Gaussian
field, and therefore gives natural extensions to the formula with
general mass functions as we have seen. Even when subdominant terms on
large scales, $Q_0$ and $Q_1$, are neglected, our result predicts that
the known result of Eq.~(\ref{eq:4-53}) should be modified to
Eq.~(\ref{eq:4-51}) with Eqs.~(\ref{eq:4-36b}) and (\ref{eq:4-36c}),
i.e.,
\begin{multline}
  \varDelta b(k) \approx
  \frac{{\sigma_M}^2}{2{\delta_{\rm c}}^2}
  \left[
    \left(2 + 2\delta_{\rm c} b^{\rm L}_1
      + {\delta_{\rm c}}^2 b^{\rm L}_2\right)\, {\cal I}(k)
  \right.
\\
  \left.
    + \left(1 + \delta_{\rm c} b^{\rm L}_1\right)
    \frac{d{\cal I}(k)}{d\ln\sigma_M}
  \right].
\label{eq:4-54}
\end{multline}
Taking the ST mass function for example, the above equation reduces to
\begin{multline}
  \varDelta b(k) \approx
   \left[
      \frac{q\delta_{\rm c} b^{\rm L}_1}{2}
      + \frac{1}{\nu^2}
       \frac{p(q\nu^2 + 2p + 1)}{1 + (q\nu^2)^p}
   \right] {\cal I}(k)
   \\
   + \left[\frac{q}{2} + \frac{1}{\nu^2}
       \frac{p}{1 + (q\nu^2)^p}\right]       
       \frac{d{\cal I}(k)}{d\ln\sigma_M}.
\label{eq:4-55}
\end{multline}
This equation is still a new result in the literature, even though
other correction terms, $Q_0$ and $Q_1$, are neglected. The previously
known result, Eq.~(\ref{eq:4-53}), is obtained only when the PS mass
function is assumed, $p=0$, $q=1$. Taking the MICE mass function, the
coefficients of Eq.~(\ref{eq:4-55}) are replaced by
Eqs.~(\ref{eq:4-43a}) and (\ref{eq:4-43b}) instead of
Eqs.~(\ref{eq:4-42a}) and (\ref{eq:4-42b}).

One should note that although the ST (MICE) mass function is derived
from Gaussian simulations, contributions of the primordial
non-Gaussianity to the mass function are higher orders in
Eq.~(\ref{eq:4-55}), and can be neglected in the lowest-order
approximation of this paper. Although higher-order corrections are
beyond the scope of this paper, they could be relevant in the
quantitative analysis of actual data.

In the specific models of primordial non-Gaussianity, large-scale
limits of the function ${\cal I}(k)$ with asymptotic bispectra of
Eqs.~(\ref{eq:3-22a})--(\ref{eq:3-22d}) are given by
\begin{align}
  {\cal I}^{\rm loc.}(k) &\approx \frac{4f_{\rm NL}}{{\cal M}(k)},
\label{eq:4-56a}\\
  {\cal I}^{\rm eql.}(k) &\approx \frac{12f_{\rm NL}}{{\cal M}(k)}
  k^2 \left[k^{2\alpha_{\rm s}} \gamma_{2+2\alpha_{\rm s}}
    - \frac{(1+\alpha_{\rm s})^2}{3} \gamma_2 \right],
\label{eq:4-56b}\\
  {\cal I}^{\rm fol.}(k) &\approx \frac{6f_{\rm NL}}{{\cal M}(k)}
  k \gamma_1,
\label{eq:4-56c}\\
  {\cal I}^{\rm ort.}(k) &\approx -\frac{12 f_{\rm NL}}{{\cal M}(k)}
  k \gamma_1,
\label{eq:4-56d}
\end{align}
where
\begin{equation}
  \gamma_\alpha(M) = \frac{1}{{\sigma_M}^2}
  \int \frac{d^3k}{(2\pi)^3} k^{-\alpha} W^2(kR) P_{\rm L}(k).
\label{eq:4-57}
\end{equation}
Predictions of the scale-dependent bias in these specific models are made
when the above equations are substituted in Eq.~(\ref{eq:4-55}). In
the scale-free power spectrum, $n_{\rm s}=1$ ($\alpha_{\rm s}=0$),
Eq.~(\ref{eq:4-56b}) is simply given by
\begin{equation}
  {\cal I}^{\rm eql.}(k) \approx \frac{8f_{\rm NL}}{{\cal M}(k)}
  k^2 \gamma_2, \quad (n_{\rm s}=1).
\label{eq:4-58}
\end{equation}

For highly biased objects, $A_2 \approx q\nu^2\delta_{\rm c}b^{\rm
  L}_1 \gg A_1$, the dominant term of Eq.~(\ref{eq:4-55}) is given by
\begin{equation}
  \varDelta b(k) \approx
      \frac{q\delta_{\rm c} b^{\rm L}_1}{2}{\cal I}(k),
\label{eq:4-59}
\end{equation}
which agrees with previous predictions \cite{dal08,MV08,SK10} in the
case of PS mass function, $q=1$. In the case of ST mass function, we
find the factor $q \simeq 0.707$ should additionally be present. Since
the ST mass function gives better fit to the halo mass function in
numerical simulations, our result suggests that these previous
predictions overestimate the amplitude of scale-dependent bias in the
high-peak limit. In the case of MICE mass function, the coefficient $q$
in Eq.~(\ref{eq:4-59}) is replaced by $2c/{\delta_{\rm c}}^2 =
0.729(1+z)^{-0.024}$.

In fact, recent numerical simulations actually prefer that the
previous theoretical predictions overestimate the amplitude of
scale-dependent bias \cite{DSI09,gro09,PPH10,WVB10,WV11}. When the
parameter $q$ in Eq.~(\ref{eq:4-59}) is freely fit to numerical
simulations, they found $q=0.6-1$ although the value varies from
simulation to simulation, depending on the algorithm to identify the
halos, etc. However, it is encouraging that the value is not so
different from $q=0.707$. It is also natural that numerical
simulations do not fit well to the high-peak formula of
Eq.~(\ref{eq:4-59}), because the correction terms in
Eq.~(\ref{eq:4-55}) are not negligible for relevant ranges of halo
mass. More quantitative comparisons of the newly derived
Eq.~(\ref{eq:4-54}) with numerical simulations, together with
estimating higher-order corrections and improving nonlocal bias
models, are left for future work.

\section{\label{sec:Numerical}
Numerical Comparisons}

In this subsection, we numerically evaluate the analytical results
derived above, and compare them with other approximate methods. For
this purpose, we use the halo bias as a specific example, where the
bias functions are given by Eqs.~(\ref{eq:4-37a}) and (\ref{eq:4-37b})
up to second order. We assume the halo sample has a redshift of $z=1$,
as a typical example.

For numerical integrations, it is useful to rewrite the function $Q_0$
of Eqs.~(\ref{eq:2-8}) and (\ref{eq:2-9a}), and the functions ${\cal
  I}_1(k)$ and ${\cal I}_2(k)$ of
Eqs.~(\ref{eq:4-38})--(\ref{eq:4-39b}) into the following
two-dimensional integrals:
\begin{align}
  Q_0(k) &= \frac{k^3}{4\pi^2} \int_0^\infty dr \int_{-1}^1 dx
  \left[2rx - \frac47 \frac{r^2(1-x^2)}{1+r^2-2rx}\right]
\nonumber\\
  &\hspace{6pc} \times
   B_{\rm L}\left(k,kr,k\sqrt{1+r^2-2rx}\right),
\label{eq:5-1a}\\
  {\cal I}_1(k) &= \frac{k^3}{2\pi^2} \int_0^\infty dr \int_{-1}^1 dx
  \,rx W\left(kR\sqrt{1+r^2-2rx}\right)
\nonumber\\
  &\hspace{6pc} \times
   B_{\rm L}\left(k,kr,k\sqrt{1+r^2-2rx}\right),
\label{eq:5-1b}\\
  {\cal I}_2(k) &= \frac{k^3}{4\pi^2} \int_0^\infty dr \int_{-1}^1 dx
  \,r^2 W(krR) W\left(kR\sqrt{1+r^2-2rx}\right)
\nonumber\\
  &\hspace{6pc} \times
   B_{\rm L}\left(k,kr,k\sqrt{1+r^2-2rx}\right).
\label{eq:5-1c}
\end{align}
In terms of the functions ${\cal I}_1(k)$ and ${\cal I}_2(k)$, the
functions $Q_1(k)$ and $Q_2(k)$ are given by Eqs.~(\ref{eq:4-39-1a})
and (\ref{eq:4-39-1b}). In numerically calculating the derivatives
$\partial {\cal I}_n/\partial\ln\sigma_M$, it is useful to note that
the derivative of the top-hat window function [Eq.~(\ref{eq:4-3-1})]
is given by
\begin{equation}
    \frac{dW(x)}{dx} = -\frac{3j_2(x)}{x}
    = \frac{3(x^2-3)\sin x + 9x \cos x}{x^4},
\label{eq:5-3}
\end{equation}
where $j_2(x)$ is the second-order spherical Bessel function. 

\begin{figure}
\begin{center}
\includegraphics[width=20.5pc]{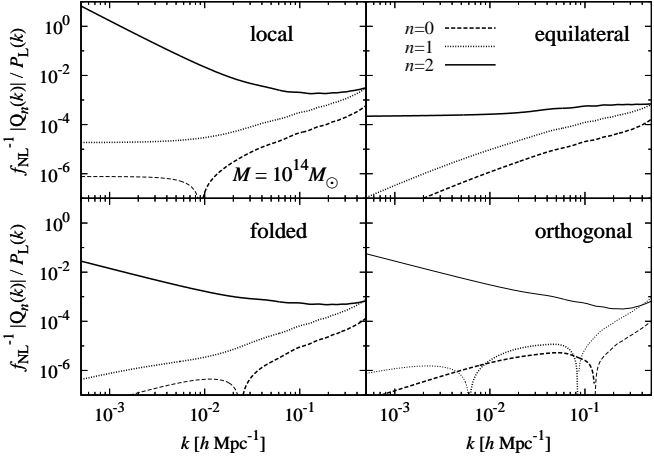}
\caption{\label{fig:Qm114} The functions $Q_n(k)$ divided by $f_{\rm
    NL} P_{\rm L}(k)$ (dashed: $n=0$, dotted: $n=1$, solid: $n=2$) at
  redshift $z=1$. Negative values are shown in thin lines. The halo
  model with mass $M=10^{14}M_\odot$ is assumed in calculating
  $Q_1(k)$ and $Q_2(k)$. Different panels correspond to different
  models of primordial non-Gaussianity as indicated.}
\end{center}
\end{figure}
In Fig.~\ref{fig:Qm114}, the functions $Q_n(k)$ ($n=0,1,2$) are
plotted in the case of $z=1$ and $M=10^{14}M_\odot$. We assume four
models of primordial non-Gaussianity,
Eqs.~(\ref{eq:3-5})--(\ref{eq:3-8}), and bias functions with ST mass
function, Eqs.~(\ref{eq:4-13a}), (\ref{eq:4-13b}), (\ref{eq:4-37a})
and (\ref{eq:4-37b}). As we have seen in
Sec.~\ref{subsec:SqueezedLimit}, the function $Q_2(k)$ is dominant over
$Q_0(k)$ and $Q_1(k)$ on sufficiently large scales.
\begin{figure}
\begin{center}
\includegraphics[width=20.5pc]{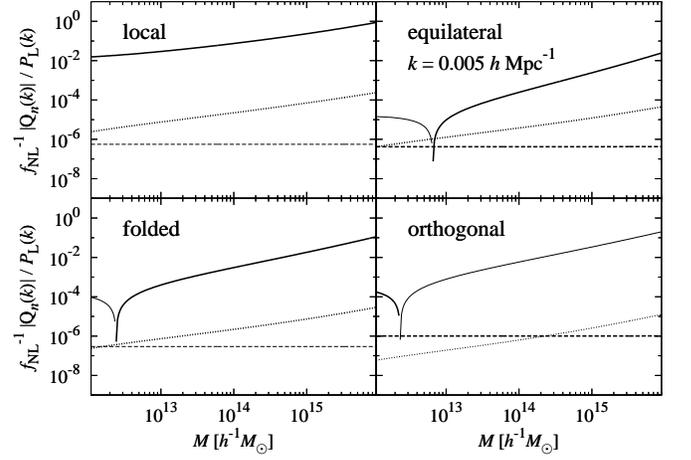}
\caption{\label{fig:Qk005} The functions $Q_n(k)$ divided by $f_{\rm
    NL} P_{\rm L}(k)$ with a fixed scale $k=0.005 \,h\, {\rm
    Mpc}^{-1}$ at redshift $z=1$, plotted against the halo mass
  (dashed: $n=0$, dotted: $n=1$, solid: $n=2$). Negative values are
  shown in thin lines. Different panels correspond to different models
  of primordial non-Gaussianity as indicated. }
\end{center}
\end{figure}
In Fig.~\ref{fig:Qk005}, the same functions are plotted against the
halo mass with a fixed scale $k=0.005\,h\,{\rm Mpc}^{-1}$. On large
scales, the function $Q_2(k)$ is dominant as expected, and the
scale-dependent bias is approximately given by Eq.~(\ref{eq:3-20}).

\begin{figure}
\begin{center}
\includegraphics[width=20.5pc]{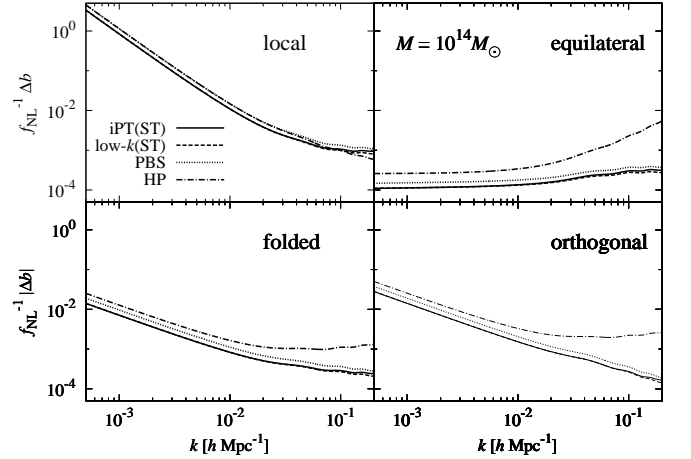}
\caption{\label{fig:m114} Scale-dependent bias $\varDelta b(k)$
  divided by $f_{\rm NL}$ for halos of mass $M=10^{14}M_\odot$. Solid
  lines correspond to predictions of iPT with ST mass function. Other
  lines correspond to predictions of the peak-background split (PBS,
  dotted lines), the low-$k$ limit of iPT with ST mass function
  (dashed lines), and the high-peak limit of PBS (dot-dashed lines).
  Positive and negative values are represented by thick and thin
  lines, respectively. When we use the PS mass function for the iPT,
  low-$k$ limit of iPT and PBS gives the same results.}
\end{center}
\end{figure}
In Fig.~\ref{fig:m114}, the scale-dependent bias $\varDelta b(k)$
divided by $f_{\rm NL}$ is plotted for $M=10^{14} M_{\odot}$ with four
models of primordial non-Gaussianity. In solid lines, the predictions
of iPT with ST mass function [Eqs.~(\ref{eq:2-13}), (\ref{eq:4-37a}),
(\ref{eq:4-37b}), (\ref{eq:4-13a}) and (\ref{eq:4-13b})] are plotted.
The predictions of the peak-background split (PBS)
[Eq.~(\ref{eq:4-53})] are plotted in dotted lines. In dashed lines,
the predictions of the large-scale (low-$k$) limit of iPT with ST mass
function [Eq.~(\ref{eq:4-55})] are plotted. In dot-dashed lines,
predictions of the PBS with high-peak and large-scale limits are
plotted. The last predictions are given by
\begin{equation}
  \varDelta b^{\rm HP}(k)
  = \frac{f_{\rm NL}\delta_{\rm c}b^{\rm L}_1}{{\cal M}(k)} \times
  \begin{cases}
      2, & \mbox{(local)}, \\
      4 k^2 \gamma_2 & \mbox{(equilateral)}, \\
      3 k \gamma_1 & \mbox{(folded)}, \\
      -6 k \gamma_1 & \mbox{(orthogonal)},
  \end{cases}
\label{eq:5-4}
\end{equation}
which are derived from Eqs.~(\ref{eq:4-56a}), (\ref{eq:4-58}),
(\ref{eq:4-56c}), (\ref{eq:4-56d}), and the first term of
Eq.~(\ref{eq:4-53}). For the equilateral model, we use the
approximation $n_{\rm s}=1$ for simplicity, and adopt
Eq.~(\ref{eq:4-58}) instead of Eq.~(\ref{eq:4-56b}). The first case of
Eq.~(\ref{eq:5-4}) is exactly the same as the original formula for the
local model derived in Refs.~\cite{dal08,slo08}. The third and forth
cases exactly match the results of Ref.~\cite{SK10} for the folded and
orthogonal models. For the equilateral model, however, the prefactor 4
is replaced by 6 in Ref.~\cite{SK10}, because angular-dependences of
the bispectrum are neglected in the latter. The factor 4 is more
accurate as shown below.
\begin{figure}
\begin{center}
\includegraphics[width=20.5pc]{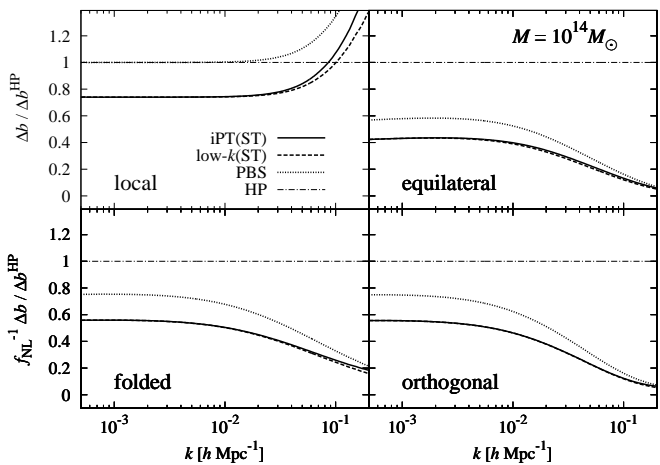}
\caption{\label{fig:m114rel} The same as Fig.~\ref{fig:m114}, but the
  scale-dependent bias $\varDelta b(k)$ is normalized by the
  large-scale, high-peak approximation $\varDelta b^{\rm HP}(k)$.}
\end{center}
\end{figure}
In Fig.~\ref{fig:m114rel}, relative amplitudes with respect to
$\varDelta b^{\rm HP}(k)$ of Eq.~(\ref{eq:5-4}) are plotted.

As expected, the low-$k$ limit of iPT (using only $Q_2$) is an
extremely good approximation on large scales ($k \alt 0.05\, h/{\rm
  Mpc}$). The large-scale asymptotes of iPT with ST mass function is
smaller than those of peak-background split. These differences
originate from the different mass functions. If we adopt PS mass
function in iPT, the asymptotes of iPT and PBS exactly agree.

The simple predictions of $\varDelta b^{\rm HP}(k)$ have correct
slopes in large-scale limits. The amplitudes are not so accurate for
non-local-type non-Gaussianities, even if we assume PS mass function
instead of ST mass function. On smaller scales ($k \agt 0.01\,h/{\rm
  Mpc}$), the slopes of $\varDelta b^{\rm HP}(k)$ are not accurate
enough. The deviations from the peak-background split in $k \agt 0.01
h^{-1}$Mpc are also pointed out in earlier work in the high-peak limit
of thresholded regions \cite{MV08,GY11}. However, higher-order loop
corrections which we do not consider in this paper might affect the
behaviors of scale-dependent bias on scales of $k \agt 0.05
h^{-1}$Mpc.

\begin{figure}
\begin{center}
\includegraphics[width=20.5pc]{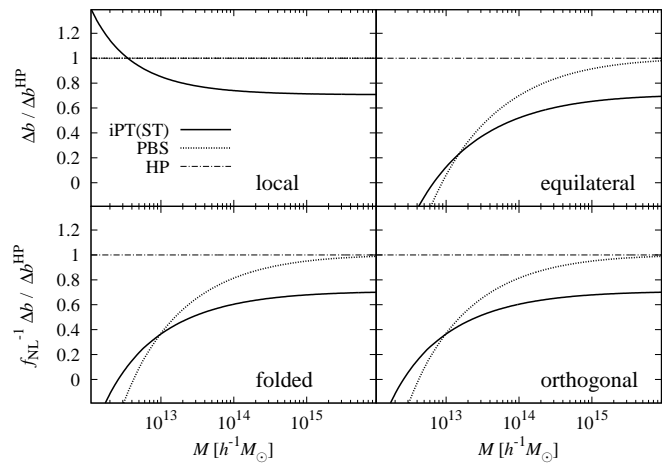}
\caption{\label{fig:k000} The normalized scale-dependent bias in
  large-scale limits ($k\rightarrow 0$, however, we numerically adopt
  $k = 10^{-4}\,h/{\rm Mpc}$ to plot this figure). As in
  Fig.~\ref{fig:m114rel}, the values are normalized by the simple
  approximation $\varDelta b^{\rm HP}(k)$. Meanings of different types
  of lines are the same as in Figs.~\ref{fig:m114} and
  \ref{fig:m114rel}. }
\end{center}
\end{figure}
In Fig.~\ref{fig:k000}, the normalized scale-dependent bias against
the halo mass is plotted in the large-scale limits ($k\rightarrow 0$).
The purpose of this figure is to illustrate the mass dependence of the
plateaux on large scales in Fig.~\ref{fig:m114rel}. The predictions of
peak-background split and high-peak limit are asymptotically agree
with each other for high-mass halos, as they should. In large-mass
limits, the predictions of peak-background split and iPT are different
by a factor of $q\simeq 0.7$, which is explained by
Eq.~(\ref{eq:4-59}).

\section{\label{sec:SDBiasR} 
Scale-dependent bias in redshift space}

Generalizing the results obtained so far in real space to those in
redshift space is fairly straightforward in the framework of iPT. The
redshift-space distortions of scale-dependent bias in the presence of
local-type non-Gaussianity have been investigated in the literature
\cite{LDS10,sch10,JSH12}, relying on the high-peak limit or the
peak-background split. With the iPT, it is straightforward to
generalize those results without restricting to the local-type
non-Gaussianity, and we do not need to rely on the high-peak limit or
the peak-background split.

When the weak time dependences in the perturbation kernels are
neglected, the $n$th order perturbations are approximately
proportional to $D^n$, and the redshift-space counterpart of the
kernel $\bm{L}_n$ is given by \cite{mat08a}
\begin{equation}
  \bm{L}^{\rm s}_n = R^{(n)}\bm{L}_n,
\label{eq:6-1}
\end{equation}
where $R^{(n)}$ is a $3\times 3$ matrix with elements
\begin{equation}
  R^{(n)}_{ij} = \delta_{ij} + nf\hat{z}_i\hat{z}_j,
\label{eq:6-2}
\end{equation}
where $f = d\ln D/d\ln a = \dot{D}/HD$ is the linear growth rate, and
$\hat{z}_i$ is the $i$th element of the unit vector $\hat{\bm{z}}$
along the line of sight. The perturbation kernels $\bm{L}_n$ in the
multipoint propagators of Eqs.~(\ref{eq:2-4-1a}) and (\ref{eq:2-4-1b})
are replaced by $\bm{L}^{\rm s}_n$ in redshift space. Specifically we
have
\begin{equation}
  \bm{k}\cdot\bm{L}^{\rm s}_n
  = (\bm{k} + nf\mu k \hat{\bm{z}})\cdot\bm{L}_n,
\label{eq:6-3}
\end{equation}
where $\mu = \bm{k}\cdot\hat{\bm{z}}/k$ is the direction cosine of the
wavevector $\bm{k}$ with respect to the line of sight.

Substituting the resulting multipoint propagators in redshift space
into Eq.~(\ref{eq:2-5b}), we have an expression for the non-Gaussian
part of the power spectrum in redshift space. In evaluating the
expression, the following integrals are useful:
\begin{align}
& \int \frac{d^3k'}{(2\pi)^3}
  \bm{L}_2(\bm{k}',\bm{k}-\bm{k}') B_{\rm L}(k,k',|\bm{k}-\bm{k}'|)
  = \frac37 \frac{\bm{k}}{k^2} R_2(k),
\label{eq:6-4a}\\
& \int \frac{d^3k'}{(2\pi)^3}
  L_{1i}(\bm{k}')L_{1j}(\bm{k}-\bm{k}')
  B_{\rm L}(k,k',|\bm{k}-\bm{k}'|)
\nonumber\\
& \hspace{4.5pc}  = \frac{k_ik_j}{k^4} \left[2 R_1(k)
      - \frac12 R_2(k)\right]
  - \frac{\delta_{ij}}{2k^2} R_2(k),
\label{eq:6-4b}\\
& \int \frac{d^3k'}{(2\pi)^3}
  c^{\rm L}_1(\bm{k}')\bm{L}_1(\bm{k}-\bm{k}')
  B_{\rm L}(k,k',|\bm{k}-\bm{k}'|)
  = \frac12 \frac{\bm{k}}{k^2} Q_1(k),
\label{eq:6-4c}\\
& \int \frac{d^3k'}{(2\pi)^3}
  c^{\rm L}_2(\bm{k}',\bm{k}-\bm{k}')
  B_{\rm L}(k,k',|\bm{k}-\bm{k}'|)
  = Q_2(k),
\label{eq:6-4d}
\end{align}
where the functions $Q_1(k)$ and $Q_2(k)$ are defined by
Eqs.~(\ref{eq:2-8}), (\ref{eq:2-9a}), (\ref{eq:2-9b}), and the
functions $R_1(k)$ and $R_2(k)$ are defined by
\begin{equation}
  R_n(k) = \int \frac{d^3k'}{(2\pi)^3}
  \hat{R}_n(\bm{k},\bm{k}') B_{\rm L}(k,k',|\bm{k}-\bm{k'}|),
\label{eq:6-5}
\end{equation}
where
\begin{align}
  \hat{R}_1 &=
  \frac{\bm{k}\cdot\bm{k}'}{k'^2},
\label{eq:6-6a}\\
  \hat{R}_2 &=
  \frac{k^2k'^2 - (\bm{k}\cdot\bm{k}')^2}{k'^2|\bm{k}-\bm{k}'|^2}.
\label{eq:6-6b}
\end{align}
Eqs.~(\ref{eq:6-4a})--(\ref{eq:6-4d}) are derived by noting that
integrals on the left-hand sides are functions of only $\bm{k}$, and
using the rotational covariance (similar technique is used in
Ref.~\cite{mat08a}). The function $Q_0(k)$ is related by
\begin{equation}
  Q_0(k) = 2R_1(k) - \frac47 R_2(k).
\label{eq:6-7}
\end{equation}
The functions $R_1(k)$ and $R_2(k)$ reduce to two-dimensional
integrals as
\begin{align}
  R_1(k) &= \frac{k^3}{4\pi^2} \int_0^\infty dr \int_{-1}^1 dx
  \,rx\,
 B_{\rm L}\left(k,kr,k\sqrt{1+r^2-2rx}\right),
\label{eq:6-8a}\\
  R_2(k) &= \frac{k^3}{4\pi^2} \int_0^\infty dr \int_{-1}^1 dx
 \frac{r^2(1-x^2)}{1+r^2-2rx}
\nonumber\\
  & \hspace{6pc}
  \times B_{\rm L}\left(k,kr,k\sqrt{1+r^2-2rx}\right).
\label{eq:6-8b}
\end{align}

Eventually, the non-Gaussian part of the power spectrum in redshift
space is given by
\begin{multline}
  \varDelta P^{\rm s}_{\rm X}(k,\mu) =
  \left[b_1(k) + f\mu^2\right]
  \Biggl\{
    2(1+f\mu^2)^2 R_1(k)
\\
    - \left[\frac47 (1+2f\mu^2) + \frac12 f^2
        \mu^2(\mu^2+1)\right] R_2(k)
\\      + (1+f\mu^2) Q_1(k)
      + Q_2(k)
  \Biggr\}.
\label{eq:6-9}
\end{multline}
When we put $f=0$ in the above equation, the power spectrum in real
space, Eq.~(\ref{eq:2-6b}), is recovered, In the large-scale limit
$k\rightarrow 0$, we have $R_1(k),R_2(k),Q_1(k) \ll Q_2(k)$, and the
asymptotic form of the power spectrum is given by
\begin{equation}
  \varDelta P_{\rm X}(k,\mu)
  \approx
  \left[b_1(k)+f\mu^2\right] Q_2(k).
\label{eq:6-10}
\end{equation}

The angular dependence of the power spectrum is conveniently
decomposed into multipoles as
\begin{equation}
  \varDelta P^{\rm s}_{\rm X}(k,\mu)
  = \sum_{l=0}^\infty \varDelta p_l(k) P_l(\mu),
\label{eq:6-11}
\end{equation}
where $P_l(\mu)$ are the Legendre polynomials, and $\varDelta p_l(k)$
are the multipole moments. Using the orthogonal relations of the
Legendre polynomials, we have
\begin{equation}
  \varDelta p_l(k) = \frac{2l+1}{2}\int_{-1}^1
  d\mu P_l(\mu) \varDelta P^{\rm s}_{\rm X}(k,\mu).
\label{eq:6-12}
\end{equation}
The monopole component $\varDelta p_0(k)$ corresponds to the angular
average of the power spectrum in redshift space, and higher-order
moments characterize anisotropies in the power spectrum, relative to
the line of sight. We decompose each multipole moment as $p_l(k) =
p^{\rm G}_l(k) + \varDelta p_l(k)$, where $p^{\rm G}_l(k)$ is the
multipole moment of the Gaussian part of the power spectrum $P^{\rm
  sG}_{\rm X}(k,\mu)$.

The multipole decomposition of the linear power spectrum in redshift
space $P^{\rm G}(k,\mu) = [b_1(k)+ f\mu^2]^2P_{\rm L}(k)$ is well known
\cite{kai87,ham98}:
\begin{align}
  p^{\rm G}_0(k) &= \left\{\left[b_1(k)\right]^2
      + \frac23 b_1(k) f + \frac15 f^2 \right\}
  P_{\rm L}(k),
\label{eq:6-13a}\\
  p^{\rm G}_2(k) &= \left\{\frac43 b_1(k) f
      + \frac47 f^2 \right\} P_{\rm L}(k),
\label{eq:6-13b}\\
  p^{\rm G}_4(k) &= \frac{8}{35} f^2 P_{\rm L}(k).
\label{eq:6-13c}
\end{align}
For the multipole decomposition of the non-Gaussian part,
Eq.~(\ref{eq:6-12}), we have
%\begin{widetext}
\begin{align}
  \varDelta p_0(k) &= 2\left[\frac{f}{3} + \frac{2f^2}{5} + \frac{f^3}{7}
      + \left(1 + \frac{2f}{3} + \frac{f^2}{5} \right) b_1(k)
  \right] R_1(k)
\nonumber\\
&  -2 \left[ \frac{2f}{21} + \frac{4f^2}{35} + \frac{3f^3}{35}
      + \left(\frac27 + \frac{4f}{21} + \frac{2f^2}{15}\right) b_1(k)
          \right] R_2(k)
\nonumber\\
&  + \left[\frac{f}{3} + \frac{f^2}{5}
      + \left(1 + \frac{f}{3}\right) b_1(k)\right] Q_1(k)
\nonumber\\
&  + \left[\frac{f}{3} + b_1(k) \right] Q_2(k),
\label{eq:6-14a}\\
  \varDelta p_2(k) &=
  4f \left[\frac13 +\frac{4f}{7} + \frac{5f^2}{21}
      + \left(\frac23 + \frac{2f}{7}
      \right) b_1(k) \right] R_1(k)
\nonumber\\
&  -f \left[ \frac{8}{21} + \frac{32f}{49} + \frac{11f^2}{21}
      + \left(\frac{16}{21} + \frac{13f}{21} \right)
      b_1(k) \right] R_2(k)
\nonumber\\
&  + 2f \left[\frac13 + \frac{2f}{7} + \frac13 b_1(k)
  \right] Q_1(k)
  + \frac{2f}{3} Q_2(k),
\label{eq:6-14b}\\
  \varDelta p_4(k) &=
  16f^2 \left[\frac{2}{35} + \frac{3f}{77} + \frac{1}{35} b_1(k)
  \right] R_1(k)
\nonumber\\
&  -\frac{4f^2}{35} \left[\frac{16}{7} + \frac{26f}{11}
      + b_1(k) \right] R_2(k)
  + \frac{8f^2}{35} Q_1(k),
\label{eq:6-14c}\\
  \varDelta p_6(k) &=
  \frac{32f^3}{231} R_1(k)
  - \frac{8f^3}{231} R_2(k).
\label{eq:6-14d}
\end{align}
%\end{widetext}
Although the above expressions are messy, most of the terms are
negligibly small in the large-scale limit $k \rightarrow 0$, and only
terms proportional to $Q_2(k)$ are dominant, i.e.,
\begin{align}
&  \varDelta p_0(k) \approx \left(\frac{f}{3} + b_1\right)
  Q_2(k),
  \quad
  \varDelta p_2(k) \approx \frac{2f}{3}Q_2(k),
\label{eq:6-15a}\\
&  \varDelta p_4(k), \varDelta p_6(k)
  \ll \varDelta p_0(k),\varDelta p_2(k),
\label{eq:6-15b}
\end{align}
where the linear (Gaussian) bias parameter $b_1$ is generally
scale-independent in the large-scale limit. The factor $f/3+b_1$ in
the monopole component is previously derived in a special case of the
local-type non-Gaussianity with the high-peak limit \cite{JSH12}.

When the bias is large enough, $c^{\rm L}_2 \gg c^{\rm L}_1 \gg 1$,
and $Q_2 \gg Q_1 \gg R_1, R_2$, we have
\begin{align}
  p^{\rm G}_0(k) &\approx [b_1(k)]^2 P_{\rm L}(k)
  \gg p^{\rm G}_2(k) \gg p^{\rm G}_4,
\label{eq:6-16a}\\
  \varDelta p_0(k) &\approx b_1(k) Q_2(k)
  \gg \varDelta p_2(k) \gg \varDelta p_4(k) \gg \varDelta p_6(k).
\label{eq:6-16b}
\end{align}
The above equations show that the redshift-space clustering reduces to
the real-space clustering when the bias is large enough. This fact is
naturally expected because the peculiar velocities of high peaks or
high-mass halos are sufficiently small.

\begin{figure}
\begin{center}
\includegraphics[width=20.5pc]{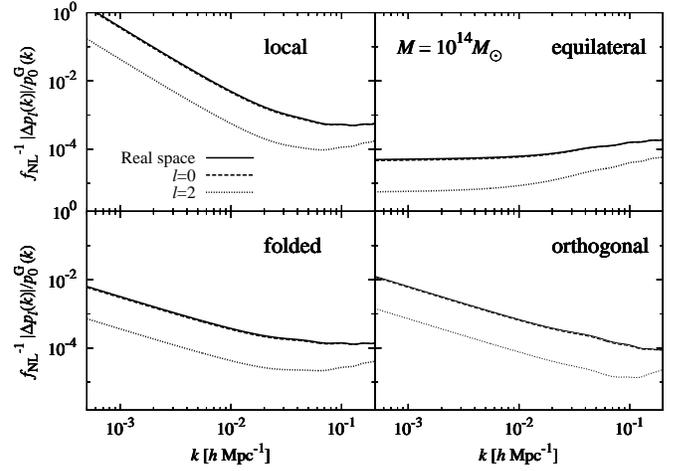}
\caption{\label{fig:rm114} Relative differences of the multipoles in
  the halo power spectrum $\varDelta p_l(k)/p^{\rm G}_l(k)$ with
  $M=10^{14}M_\odot$ (dashed lines: $l=0$, dotted lines: $l=2$). Solid
  lines corresponds to the relative differences of the real-space
  power spectrum, which are the same as the solid lines in
  Fig.~\ref{fig:m114}. Positive and negative values
  are represented by thick and thin lines, respectively. }
\end{center}
\end{figure}
In Fig.~\ref{fig:rm114}, relative differences of the lower-order
multipoles, $\varDelta p_l(k)/p^{\rm G}_0(k)$, are plotted for
$M=10^{14} M_\odot$. We only plot the lowest two multipoles $l=0,2$,
since the higher-order multipoles $l=4,6$ are small enough. The scale
dependences of multipoles with $l=0,2$ have a similar slope on large
scales but different amplitudes. These properties are explained by the
dominant contributions on large scales, Eq.~(\ref{eq:6-15a}). The
ratio of those two multipoles, in the large-scale limit, is given by
\begin{equation}
  \frac{\varDelta p_2(k)}{\varDelta p_0(k)}
  \approx
  \frac{2f}{f+3b_1} = \frac{2}{1+3/\beta},
\label{eq:6-17}
\end{equation}
where $\beta \equiv f/b_1$ is the linear redshift-space distortion
parameter. This ratio is small for highly biased objects, or massive
halos.

Relative monopole components $\varDelta p_0(k)/p^{\rm G}_0(k)$ on
large scales are not so different between real space and
redshift-space. This property is true even in the low-mass halos.
Therefore, the scale-dependent bias is not so affected by the
redshift-space distortions when only the monopole component is
concerned. However, the quadrupole component may be used in
constraining the primordial non-Gaussianity.

\begin{figure}
\begin{center}
\includegraphics[width=20.5pc]{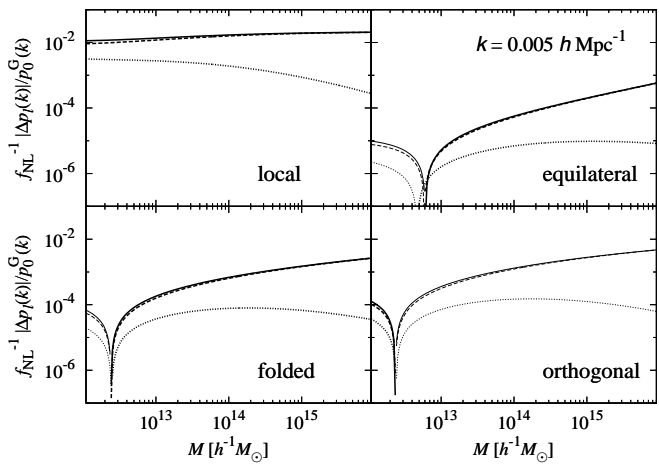}
\caption{\label{fig:rk005} Relative differences of the multipoles in
  the halo power spectrum as a function of the halo mass for a fixed
  scale $k=0.005\,h^{-1}{\rm Mpc}$. Meanings of different types of
  lines are the same as in Fig.~\ref{fig:rm114} (solid lines: real
  space, dashed lines: $l=0$, dotted lines: $l=2$). }
\end{center}
\end{figure}
In Fig.~\ref{fig:rk005}, relative differences of the lower-order
multipoles are plotted against the mass of halo, where the scale is
fixed to $k=0.005\,h^{-1}{\rm Mpc}$. As described above, the relative
monopole components are similar in real and redshift spaces,
irrespective to the mass of halos. They are asymptotically the same in
the high-mass limit, because of Eqs.~(\ref{eq:6-16a}) and
(\ref{eq:6-16b}). In the same limit, the ratio $\varDelta
p_2/\varDelta p_0$ approaches to $2\beta/3$ according to
Eq.~(\ref{eq:6-17}).

\section{Conclusions
\label{sec:concl}
}

The iPT is a general framework of the perturbation theory in the
presence of bias. In this paper, we first apply this framework to
deriving the relation between the scale-dependent bias and the
primordial non-Gaussianity. Approximations such as the peak-background
split and the high-peak limit, which are usually adopted in the literature
to estimate the scale-dependent bias, are not required. The
redshift-space distortions of the scale-dependent bias are also
evaluated. Thus, in this paper, we have derived the most general
formula so far of the scale-dependent bias with primordial
non-Gaussianity in the literature.

For the scale-dependent bias in real space, the most fundamental
equation in this paper is provided by Eq.~(\ref{eq:2-13}), where
$Q_n(k)$ is linearly dependent on the primordial bispectrum. We find
that the slope of the scale-dependent bias in the large-scale limit is
determined only by primordial bispectra in the squeezed limit, and is
independent on detailed models of bias. This property explains the
fact that different models of bias have predicted the same slope of
the scale-dependent bias in the literature.

We derive the shape of renormalized bias functions, generalizing the
concept of simple Press-Schechter approach. The general expression of
renormalized bias functions in this approach is given by
Eq.~(\ref{eq:4-26}). In the case of universal mass function, the
renormalized bias functions are given by Eq.~(\ref{eq:4-32}), or
equivalently Eq.~(\ref{eq:4-33}). The previously known results in the
approximation of peak-background split are reproduced when the PS mass
function is assumed in our results [Eq.~(\ref{eq:4-53})]. When the
mass function deviates from the PS form, our results suggest that the
form of scale-dependent bias should be corrected. The general formula
of scale-dependent bias on large scales is given by
Eq.~(\ref{eq:4-54}). This equation is one of the main outcomes in this
paper. Most of the previous results regarding the scale-dependent bias
from the primordial non-Gaussianity are derived as special cases of
this equation.

The evaluations of the redshift-space distortions in the
scale-dependent bias are straightforward in the framework of iPT. The
result is given by Eq.~(\ref{eq:6-9}), or in terms of multipole
coefficients, Eqs.~(\ref{eq:6-14a})--(\ref{eq:6-14d}). On large
scales, however, dominant terms in these equations are simply given by
Eqs.~(\ref{eq:6-15a}) and (\ref{eq:6-15b}). When the bias is large
enough, the redshift-space distortions do not affect the
scale-dependence of bias much [Eqs.~(\ref{eq:6-16a}) and
(\ref{eq:6-16b})]. Even when the bias is not large enough, the
redshift-space distortions have little effects at least in the
non-Gaussian models we have considered (Figs.~\ref{fig:rm114}
and \ref{fig:rk005}).

While highly biased objects have large amplitudes of power spectrum,
the number of objects is small and the shot noise is large. Thus
highly biased objects are not suitable for testing the primordial
non-Gaussianity. On the other hand, the amplitude of power spectrum is
small for less biased objects, and the clustering signals are small.
Consequently, there should be an optimal objects with sufficiently
large bias and sufficiently large numbers at the same time for
realistically constraining the primordial non-Gaussianity by galaxy
surveys. The high-peak limit or the peak-background split are not
necessarily valid in some cases. The results of this paper provide the
most accurate formula of the scale-dependent bias in the literature.
They should be useful in theoretical investigations as well as in
constraining the primordial non-Gaussianity with realistic galaxy
surveys. Applications of the results in this paper, including the
Fisher analysis of the future galaxy surveys, higher-order analyses of
primordial non-Gaussianity, are now in progress. For more accurate
modeling of the scale-dependent bias, it should be necessary to
improve the nonlocal bias model beyond the simple halo approach.
Investigations in this direction will be addressed in future work.

\begin{acknowledgments}
    I wish to thank S.~Yokoyama and V.~Desjacques for helpful
    discussion. I acknowledge support from the Ministry of Education,
    Culture, Sports, Science, and Technology, Grant-in-Aid for
    Scientific Research (C), 24540267, 2012, and Grant-in-Aid for
    Scientific Research on Priority Areas No. 467 ``Probing the Dark
    Energy through an Extremely Wide and Deep Survey with Subaru
    Telescope.'' This work is supported in part by JSPS (Japan Society
    for Promotion of Science) Core-to-Core Program ``International
    Research Network for Dark Energy.''
\end{acknowledgments}

%%%%%%%%%%%%
\renewcommand{\apj}{Astrophys. J.}
\newcommand{\aap}{Astron. Astrophys. }
\newcommand{\apjl}{Astrophys. J. Letters }
\newcommand{\apjs}{Astrophys. J. Suppl. Ser. }
\newcommand{\apss}{Astrophys. Space Sci. }
\newcommand{\jcap}{J. Cosmol. Astropart. Phys. }
\newcommand{\mnras}{Mon. Not. R. Astron. Soc. }
\newcommand{\mpla}{Mod. Phys. Lett. A }
\newcommand{\pasj}{Publ. Astron. Soc. Japan }
\newcommand{\physrep}{Phys. Rep. }
\newcommand{\ptp}{Progr. Theor. Phys. }
\newcommand{\jetp}{JETP }
%\newcommand{\prl}{Phys. Rev. Lett.}

%\bibliography{redoneloop}% Produces the bibliography via BibTeX.

\end{document}